\documentclass[12pt]{article}
\usepackage[utf8]{inputenc}
\usepackage[left=2.5cm,right=2.5cm,top=2.5cm,bottom=3cm]{geometry}
\usepackage{graphicx}
\usepackage[utf8]{inputenc}
\usepackage[T1]{fontenc}
\usepackage{lmodern}
\usepackage{microtype}
\usepackage{todonotes}
\usepackage{amsmath}
\usepackage{slashed}
\usepackage{amssymb}
\usepackage{amsfonts}
\usepackage{mathrsfs}
\usepackage{appendix}
\usepackage{mathtools}
\usepackage{tabularray}
\usepackage{bbold}
\usepackage{comment}
\usepackage{cite}
\usepackage{hyperref}
\hypersetup{linktocpage=true,colorlinks=true,citecolor=blue,linkcolor=blue,urlcolor=black}

\usepackage{tikz}
\usepackage{physics}     
\usepackage{bm}         
\makeatletter
\@addtoreset{equation}{section}
\makeatother

\newcommand{\fatdash}{\raisebox{0.4ex}{\rule{0.6em}{1.2pt}}}

\title{Couch-Torrence inversion of D3-brane backgrounds}
\author{Antonio Cristofaro}
\date{November 2025}

\DeclareMathOperator{\im}{\mathrm{i}}
\begin{document}
\bibliographystyle{mystyle}
\vspace{20pt}

\Large 
 \begin{center}
  \textbf{Couch-Torrence conformal inversion, supersymmetry and conserved charges 
  for D3-branes  
  }
\pagenumbering{gobble}
\vskip 1cm  
\large   
{Mohammad Akhond$^*$
, Massimo Bianchi$^*$, Antonio Cristofaro$^\ddagger$, Fabio Riccioni$^\dagger$}\\

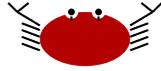
\hspace{20pt}
\hspace{20pt}

\small  
$^*$\it Sezione INFN Roma “Tor Vergata” \& Dipartimento di Fisica,
Universita di Roma “Tor Vergata”, Via della Ricerca Scientifica 1, 00133, Roma, Italy\\
$^\ddagger$ Dipartimento di Fisica, Università di Roma “La Sapienza”, \\ Piazzale Aldo Moro 2, 00185, Roma, Italy\\
$^\dagger$Sezione INFN Roma “La Sapienza”, Piazzale Aldo Moro 2, 00185, Roma, Italy\\
\end{center}
\normalsize
\vspace{5pt}
\begin{abstract}

An asymptotically flat spacetime in $D=4$ can be mapped via Couch-Torrence conformal inversion to the geometry around an extremal non-expanding and non-rotating horizon. At the linearized level, an infinite tower of conserved Newman-Penrose charges can be found at null-infinity, while infinitely many Aretakis charges are conserved in the near-horizon. Couch-Torrence inversion allows one to establish a matching between the two sets of asymptotic charges. In this work we construct the Newman-Penrose and Aretakis scalar charges in higher-dimensional geometries of D3-branes in $D=10$ and D3-brane bound states in $D=4$ and $D=5$ and establish a precise matching between them through the inversion. By exploiting the residual unbroken supersymmetry of Type-IIB supergravity, we demonstrate that it is possible to relate scalar (complex dilaton) charges to spinor and tensor charges. In particular, we determine infinite towers of conserved asymptotic spinorial charges associated with the dilatino fluctuations, and determine the map through inversion. 
\end{abstract}
\newpage
\tableofcontents
\newpage
\pagenumbering{arabic}

\section{Introduction}

As shown long ago by Couch and Torrence (CT) in their seminal paper \cite{Couch:1984}, extremal rotating and non-rotating black holes (BHs) in $D=4$ are known to admit a peculiar symmetry under conformal inversions. Later on, this symmetry was revived in the context of STU (non-)rotating BHs \cite{Cvetic:2020kwf} in supergravity and superstring theories and in connection to Aretakis charges \cite{Aretakis:2012ei, Godazgar:2017igz, Aretakis:2011ha, Godazgar:2020gqd}. More recently, one of the present authors (MB) in collaboration with Giorgio Di Russo extended the analysis to D3-branes and their bound states as well as to other branes and BHs in higher dimensions \cite{Bianchi:2021yqs, Bianchi:2022wku}. One of the main results was the role of the light-ring as fixed locus of CT inversions in the non-rotating cases. Moreover a relation was established between observables such as the deflection angle for hyperbolic-like (open) motion and infalling angle for elliptic-like (closed) motion with some similarity with the B2B relation proposed in \cite{Kalin:2019rwq} that however requires an analytic continuation.
In the case of rotating STU BHs, multi-charge rotating BHs in STU supergravity, which can be constructed as bound states of 4 stacks of wrapped D3-branes, the precise conditions for CT inversion to hold were identified and hints to a generalized holographic interpretation were given. Further work on generalized CT inversion exploiting the role of the photon ring or photon shells of non-extremal BHs in (A)dS has appeared in \cite{Charalambous:2025hlc}.

More recently, an inspiring investigation of the geometric duality  \cite{Agrawal:2025fsv} was performed that allowed to relate Newman-Penrose (NP) charges \cite{Newman:1961qr} associated with fields of spin $s=0,1,2$ at null infinity and Aretakis charges \cite{Aretakis:2011ha} at extremal non-expanding (non-) rotating horizons in $D=4$. 
For extremal (non-)rotating BHs in $D=4$, the conservation of Aretakis charges has been shown to entail linear instability of the horizon \cite{Aretakis:2012ei}. To the best of our knowledge, non-linear effects that may spoil the conservation of these charges and invalidate the instability analysis have not been studied in detail. Connection with the emergence of scale and conformal invariance at extremality has been given support recently \cite{Gralla:2017lto}. In \cite{Chen:2025sim} it was shown that also non-dilatonic extremal black $p$-branes in $D$ dimensions suffer from Aretakis instability. This is associated with the Kaluza-Klein perturbations arising from the compactification on the $S^{D-p-2}$ factor in the near-horizon geometry. This result was found using symmetry arguments \cite{Gralla:2018xzo} rather than by constructing conserved quantities.

In the present investigation we show that generalized CT investigations allow us to relate Newman-Penrose (NP) charges at null infinity and Aretakis charges at extremal non-expanding (non-)rotating horizons of D3-branes in $D=10$ and their bound states in $D=5$ and $D=4$. Moreover, we exploit the residual unbroken SUSY to relate scalar (complex dilaton) charges to spinorial and tensorial charges. To this end we (re)derive the explicit expressions of the Killing spinors and solve the dilatino field equation in order to identify the spinorial charges and relate them to the scalar charges. The latter may be related to dynamical multipoles of the radiative field at null infinity or to multipolar modes at the horizon. We conjecture that their conservation at the linear level is related to peculiar properties of the Kaluza-Klein modes of Type-IIB supergravity in AdS$_5 \times$S$^5$, thus suggesting a holographic origin for the phenomenon. On the other hand, thanks to the geometric duality between null infinity and extremal non-expanding (non-)rotating horizons put forward in \cite{Agrawal:2025fsv} it is tantalizing to speculate on a possible holographic interpretation of the conserved NP charges as KK modes of the dilaton and its superpartners also at null infinity. This observation may give more than a hint to Carrollian flat-space holography \cite{Ruzziconi:2026bix, Bergshoeff:2020xhv}, possibly in connection with celestial holography \cite{Donnay:2023mrd} and BMSVB (Bondi-Metzner-Sachs and Van de Burg) super-translations and super-rotations \cite{Bondi:1962px, Sachs:1962zza,  Barnich:2011mi, Strominger:2013jfa}. One should, however, be cautious of the classical nature of this correspondence. After all, Weyl anomalies could require corrections to the map between the two sets of asymptotic charges. This treatment goes beyond the scope of our current analysis.

The plan of the present work is as follows. In Section \ref{GenSet} we will describe the general setting that allows to extend CT conformal inversions to spacetimes with extremal non-expanding horizons and extra dimensions and branes. For illustrative purposes, we discuss scalar charges at null infinity (Newman-Penrose) and near-horizon (Aretakis) and relate them and review how to apply CT inversions to the simple self-dual case of RN BHs. \footnote{Self-duality here refers to the conformal invariance of the background under the coordinate inversion, and should not be confused with electric-magnetic duality.}

We then pass to consider D3-branes and their bound states in Section \ref{GenD3branes}. Starting with a single D3-brane stack,  we find an infinite number of scalar NP charges associated with the radiative multipoles at null infinity in the transverse six dimensions and thus labelled by a positive integer multipole number $\ell$ and its three projections $\vec{m}=(m_1, m_2, m_3)$. The 1/2 BPS nature of the configuration is mentioned but not exploited at this point. A possible holographic interpretation of the conserved Aretakis charges near-horizon in terms of KK modes of the complex dilaton on $S^5$  is offered that suggests a possible interpretation of the conserved NP charges in terms of multipolar modes of the complex dilaton at null infinity.

1/4 BPS configurations (preserving 8 supercharges in Type IIB) of two (smeared) intersecting stacks of D3-branes are studied in the simple case with $N_1=N_2$ and the radiative and near-horizon scalar charges identified and connected to each other. Configurations of four intersecting stacks of D3-branes, preserving 4 supercharges {\it i.e.} 1/8 BPS in Type IIB, are studied and related to multi-charge STU BHs. The conditions  for CT conformal symmetry are reviewed and scalar NP and Aretakis charges are identified and connected to each other.

The crucial role of supersymmetry is investigated in Section \ref{SUSY}. First, we review the explicit derivation of the 16 Killing spinors associated to 16 globally preserved supercharges for D3-branes \cite{Ortin:2015hya}. We then argue that unbroken SUSY transforms boson and fermion charges into one another. To support our observation, we explicitly construct 16 infinite towers of dilatino charges associated to spin 1/2 spherical harmonics on $S^5$. We also sketch how to extend the analysis to the other components of the Type-IIB supergravity multiplet. 

Section \ref{conclusions} contains a summary of our results and a discussion of the implications of our analysis for flat-space holography, possibly after taking into account non-linear effects, and some suggestions for future research. 

Appendix \ref{appendixA} contains our conventions, frame notation, Clifford algebras in various dimensions, and other technical details. Appendix \ref{Appendix: KK modes} contains a detailed summary of the KK mode expansion of the supergravity fields on $S^5$.

\section{General setting}\label{GenSet}
The aim of this section is to generalize the Couch-Torrence (CT) conformal inversion to self-dual, higher-dimensional backgrounds, such as D3-brane solutions and  bound states thereof. We begin with a pedagogical introduction to CT inversion using as a prototype the extremal Reissner-Nordstr{\"o}m solution, where we derive the conserved charges in both the near-infinity and near-horizon limits. We then demonstrate how the CT inversion leads to a direct matching between these two sets of asymptotic charges. Later, we extend this formalism to higher-dimensional self-dual geometries, including those with compact dimensions. Finally, we establish the multipole expansions for a scalar field at large and small radius in these backgrounds, providing the necessary framework to define NP charges and Aretakis charges in these geometries.

\subsection{Extremal Reissner-Nordstr{\"o}m: a pedagogical case study}
Consider the metric of an extremal Reissner-Nordstrom (RN) black hole (BH)
\begin{equation}
    ds^2 = - f(r) dt^2 + \frac{dr^2}{f(r)} + r^2 d\Omega^2_{2} \ ,
\end{equation}
where $f(r) = \left(1-\frac{M}{r}\right)^2$ is the red-shift factor (for $M=|Q|$) and $d\Omega^2_{2} = d\vartheta^2 + \sin^2\vartheta d\phi^2$ is the metric on $S^2$. The horizon corresponds to $r_H = M$ while the photon sphere is at $r_c = 2M$.

Shifting the radial coordinate as
\begin{equation}
    \rho = r - M \ ,
\end{equation}
the horizon and the photon sphere are shifted to $\rho_H = 0$ and $\rho_c = M$.  
One can then define the retarded and advanced time $u=t-\rho^*$ and $v=t+\rho^*$, where $\rho^*$ is the tortoise coordinate, defined such that 
\begin{equation}
    d\rho^* = \frac{(M+\rho)^2}{\rho^2}d\rho \ .
\end{equation}
In particular, choosing $\rho^*$ such that $\rho^*(\rho_c) = 0$, yields
\begin{equation}
    \rho^* = \rho - \frac{M^2}{\rho} + 2M\ln\frac{\rho}{M} \ .
\end{equation}
In outgoing Eddington-Finkelstein coordinates $\{u,\rho,\vartheta,\phi\}$ the metric reads
\begin{equation} \label{eq: rn inf}
    d  s^2_{\mathcal{I}} = -   \frac{\rho^2}{(M+\rho)^2} du^2 -2dud\rho +(M+\rho)^2 d\Omega^2_{2} \ ,
\end{equation}
while in ingoing coordinates $\{v,\rho,\vartheta,\phi\}$ it takes the form
\begin{equation} \label{eq: rn hor}
    ds^2_{\mathcal{H}}= -   \frac{\rho^2}{(M+\rho)^2} dv^2 + 2dvd\rho + (M+\rho)^2 d\Omega^2_{2}  \ .
\end{equation}

The Couch–Torrence (CT) inversion symmetry is a radial inversion, effectively exchanging the near-horizon region with null infinity, and it is realized in RN BH as the conformal map
\begin{equation}\label{eq: rs conf map}
    u \rightarrow v \qquad \rho \rightarrow  \frac{M^2}{\rho} \ ,
\end{equation}
that relates the retarded and advanced metrics in eqs. \eqref{eq: rn inf}  and \eqref{eq: rn hor}  as \footnote{Note that this form of the metric, although suited for the description of past and future null infinity, is not the metric on the conformal boundary. Therefore the emergence of Carroll geometry is not yet apparent, while it is of crucial significance for exploiting CT inversion in the Carrollian perspective on flat-space holography. We thank Luca Romano for this observation.}
\begin{equation}
    ds^2_{\mathcal{I}} \rightarrow \frac{M^2}{\rho^2}  d s^2_{\mathcal{H}} \ .
\end{equation}
Under this conformal inversion  $\rho^* \to -\rho^*$ so the  photon sphere $\rho^*=0$ is a fixed point of this symmetry and the effective potential for null geodesics or waves is an even function of $\rho^*$.

Now consider the dynamics of an outgoing massless test scalar field $\Phi$ in the background geometry of the metric \eqref{eq: rn inf}, by solving the Klein-Gordon equation
\begin{equation}
    \Box_{\mathcal{I}}\Phi = 0 
\end{equation}
through the near-infinity asymptotic expansion
\begin{equation}
    \Phi(u,\rho, \vartheta, \phi) = \sum_{n=0}^\infty \frac{1}{\rho^{n+1}}\sum_{\ell=0}^{\infty}
    \sum_{m=-\ell}^{+\ell}\varphi^{(n)}_{\ell, m}(u)Y_{\ell,m}(\vartheta, \phi) \ ,
\end{equation}
where $Y_{\ell,m}$ are the spherical harmonics on $S^2$.
Denoting the derivative with respect to $u$ by a dot and dropping an overall factor $(M+\rho)^{-2}$ ,  the Klein-Gordon equation reads
\begin{align*}
    & \sum_{\ell,m} \Big\{\sum_{n} \Big[-\ell(\ell+1) + n(n+1)\Big] \rho^{-n-1}\varphi^{(n)}_{\ell,m} + 2M^2 \sum_{n} (n+1) \rho^{-n-2}\dot \varphi^{(n)}_{\ell,m} \\
    &+ 2M \sum_{n}(2n+1) \rho^{-n-1} \dot \varphi^{(n)}_{\ell,m} + 2\sum_{n} n \rho^{-n} \dot \varphi^{(n)}_{\ell,m} \Big\} Y_{\ell,m} = 0 \ .
\end{align*}
Orthogonality of the spherical harmonics implies that each term in the bracket has to separately vanish, resulting in the equation 
\begin{equation}
     [-\ell(\ell+1) + n(n+1)] \varphi^{(n)}_{\ell,m} + 2M^2  n \dot \varphi^{(n-1)}_{\ell,m} 
    + 2M (2n+1)  \dot \varphi^{(n)}_{\ell,m} + 2 (n+1) \dot \varphi^{(n+1)}_{\ell,m}=0  
\end{equation}
for each coefficient in the $1/\rho$ expansion. Noting that the first term in the equation above vanishes for $n = \ell$, one can extract the infinite tower of conserved charges \cite{Aretakis:2011ha,Aretakis:2012ei,Godazgar:2017igz, Godazgar:2020gqd,Agrawal:2025fsv}
\begin{equation}
N_{\ell, m} =  2M^2 \ell \varphi^{(\ell-1)}_{\ell, m} + 2M(2\ell +1) \varphi^{(\ell)}_{\ell, m} + 2( \ell + 1) \varphi^{(\ell+1)}_{\ell, m} \ .
\end{equation}
These  charges  are conserved in the sense that 
\begin{equation}
    \partial_u  N_{\ell, m}=0 \ ,
\end{equation} barring non-linear terms arising from self-interactions of the scalar field or back-reaction on the metric.

Let us now consider the near-horizon expansion of the incoming massless scalar field $  \widetilde \Phi$ on the metric \eqref{eq: rn hor}. In this case the field expansion is
\begin{equation}
    \widetilde \Phi(v,\rho, \vartheta, \phi) = \sum_{n=0}^\infty \rho^n \sum_{\ell=0}^{\infty}
    \sum_{m=-\ell}^{+\ell}\widetilde\varphi^{(n)}_{\ell, m}(v)Y_{\ell,m}(\vartheta, \phi) \ .
\end{equation}
Repeating the same computation, at the end one obtains the conserved Aretakis charges
\begin{equation}
    A_{\ell,m} = 2\ell \widetilde \varphi^{(\ell-1)}_{\ell,m} + 2M(2\ell+1) \widetilde \varphi^{(\ell)}_{\ell,m} + 2M^2(\ell +1) \widetilde \varphi^{(\ell+1)}_{\ell,m} \qquad \partial_v A_{l,m}=0 .
\end{equation}

Since ERN background has vanishing scalar curvature $R=0$, the equation of motion for the scalar field is conformally invariant. This means that under the map \eqref{eq: rs conf map}  the scalar field transforms as
\begin{equation}
    \widetilde \Phi (v, \rho, \vartheta, \phi) = \frac{M}{\rho} \Phi (u \to v, \rho\to \frac{M^2}{\rho}, \vartheta, \phi)
\end{equation}
and hence
\begin{equation}
    \widetilde \varphi ^ {(n)}_{\ell,m} = \frac{1}{M^{2n +1}} \varphi^{(n)}_{\ell,m}\ .
\end{equation}
Notice that the scaling dimension is, in general, related to $n$ rather than to $\ell$.
Using this relations one can obtain the matching between Aretakis and Newman-Penrose charges:
\begin{equation}
    A_{\ell,m} = \frac{1}{M^{2\ell+1}} N_{\ell,m} \ .
    \label{eq: ANPERN}
\end{equation}
Up to a (rather trivial) mass factor they actually coincide. Since the near-infinity charges are mapped to the near-horizon charges with the same $(\ell, m) $, the conformal inversion preserves the multipolar structure of charges.

\subsection{Generalized Couch-Torrence inversion}

In order to generalize the CT inversion for extremal BHs to a single stack of $N$ D3-branes and to bound states thereof (two-stacks and four-stacks intersections), consider a $D=(\hat{p}+1) + (d+1)+ d_z$-dimensional asymptotically flat spacetime described by coordinates $X^M = (x^{\mu}, y_i, \mathbf{z})$. Here $x^\mu = (t,\vec{x})$ with $\mu = 0,..., \hat{p}$, are the  longitudinal directions common to all the stacks of D3-branes, $y_i $ denote the $d+1$ transverse directions to all the D3-brane stacks, while $\mathbf{z}$ represents possible additional compactified $d_z$ dimensions with mixed boundary conditions, corresponding to Neumann directions for one (or more) stacks and Dirichlet directions for the other(s). 

The  metric describing D3-branes and their bound states takes the general form\footnote{For simplicity we assume that all the stacks consist of the same numbers of D3-branes.}
\begin{equation}
    ds^2 = h(y_i)^{-1} \eta_{\mu \nu}dx^{\mu}dx^\nu + h(y_i) \delta_{ij}dy^idy^j + d\mathbf{z}^2 \ ,
\end{equation}
where $h(y_i)$ denotes the relevant redshift factor in the D3-brane background or its bound states, and $d\mathbf{z}^2$ is the metric associated with the compactified $d_z$ coordinates. We adopt the mostly-plus convention for the Minkowski metric $\eta_{\mu \nu}$.

\begin{table}[]
    \centering
    \begin{tabular}{c|c|c|c}
          &D3 & D3-D3$'$ & D3-D3$'$-D3$''$-D3$'''$ \\
          \hline
       $\hat{p}+1$  & $4$ & $2$  & $1$ \\  
       $d+1$ & $6$ & $4$ & $3$ \\
       $d_z$ & 0& $4$ & $6$
    \end{tabular}
    \caption{Number of longitudinal, transverse, and compactified ND directions for the D3-brane configurations considered in this work.}
    \label{table: dim branes}
\end{table}

By rotational symmetry in the transverse space, we introduce the radial coordinate $r^2=\sum_i y_i^2$ and angular coordinates $\vartheta^a$, with $a=1,\dots, d$ parametrizing the unit $d$-sphere $S^d$. 

In order to study the near-infinity region, we define the retarded time $u = t - r^*$, where $r^*$ is the tortoise coordinate defined as $dr^* = h(r)dr$. In outgoing coordinates $X^M = (u,\vec{x}, r, \vartheta^a, \mathbf{z})$, the metric takes the form
\begin{equation}
    ds^2_{\mathcal{I}} = h(r)^{-1} (-du^2 + d\vec{x}^2) - 2 du dr+ h(r) r^2 \gamma_{ab}(\vartheta^c) d\vartheta^ad\vartheta^b + ds^2_{\mathbf{z}} \ ,
\end{equation}
where $\gamma_{ab}$ denotes the metric of $S^d$.

In the near-horizon geometry we use instead ingoing coordinates $X'^M = (v, r, \vartheta^a, \vec{x}, \mathbf{z})$, where $v$ is the advanced time coordinate $v = t+r^*$. The metric is similar to the previous one
\begin{equation}
    ds^2_{\mathcal{H}} = h(r)^{-1} ( -dv^2 + d\vec{x}^2 ) + 2 dv d r + h(r) r^2 \gamma_{ab}(\vartheta^c) d\vartheta^a d \vartheta^b + ds^2_{\mathbf{z}}\ .
\end{equation}
The coordinates ${\mathbf{z}}$ will largely be spectators in the following. 

Consider the CT conformal inversion
\begin{equation}\label{eq: conf inv}
    u \to v \qquad r \to\frac{r_c^2}{r} \ ,
\end{equation}
where $r_c$ is a characteristic length scale, identified with the photon-sphere radius, while the remaining directions are inert. The D3-brane solution and its bound states are self-dual under \eqref{eq: conf inv}, {\it i.e.} the metric associated to the $(\hat{p}+1)+(d+1)$-dimensional  outgoing coordinates $X^{\tilde{M}}=(u,\vec{x},r,\theta^a)$  maps to the one associated to the  $(\hat{p}+1)+(d+1)$-dimensional ingoing coordinates $X'^{\tilde{M}}=(v,\vec{x},r,\theta^a)$, up to a Weyl rescaling. Explicitly:
\begin{equation}\label{conf dual}
    ds_{\mathcal{I}}^2 - ds^2_{\mathbf{z}} \to \alpha^2 \left( d s_{\mathcal{H}}^2 - ds^2_{\mathbf{z}} \right) \;,\qquad \alpha = \frac{r_c}{r} \ .
\end{equation}

In the following we will exploit this CT symmetry \cite{Couch:1984}, originally found in the context of extremal BHs and later on extended to multi-charge (non-)rotating BHs \cite{Cvetic:2020kwf} and, more importantly for our present purposes, generalized to (extremal) D3-branes and their bound states \cite{Bianchi:2021yqs, Bianchi:2022wku}. 

For simplicity we initially focus on identifying the scalar NP charges near null-infinity and relating them to near-horizon  Aretakis charges. We will then exploit unbroken SUSY to relate the scalar charges to spinorial and tensorial charges.

\subsection{Multipolar expansion of a scalar field}

The (radiative) multipolar expansion of a scalar field $\Phi$ for large $r$ reads
\begin{equation}
    \Phi(u,r,\vartheta^a, \vec{x}, \mathbf{z}) = \sum_{n=0}^\infty\sum_{\ell=0}^{\infty} \sum_{\vec{m}\in M_\ell}\frac{1}{r^{n+\delta}} \varphi^{(n)}_{\ell,\vec{m}}(u, \vec{x}, \mathbf{z})Y_{\ell,\vec{m}}(\vartheta^a) \ ,
\end{equation}
where $Y_{\ell,\vec{m}}$ are scalar spherical harmonics, satisfying
\begin{equation}
    \nabla^2 Y_{\ell,\vec{m}} = -\ell \left( \ell + {d-1}\right) Y_{\ell,\vec{m}}  \ ,
\end{equation}
with ${\vec{m}\in M_\ell}$ representing the independent projections (Dynkin labels) of the totally symmetric traceless tensors labelled by $\ell$. The large $r$ behaviour is controlled by the parameter
\begin{equation}
\delta ={d-1} \ .
\end{equation}

The near-horizon expansion is instead
\begin{equation}
    \widetilde \Phi (v,r,\vartheta^a, \vec{x}, \mathbf{z}) = \sum_{n=0}^\infty\sum_{\ell=0}^{\infty} \sum_{\vec{m}\in M_\ell} r^n \widetilde \varphi ^{(n)}_{\ell,\vec{m}}(v, \vec{x}, \mathbf{z}) Y_{\ell,\vec{m}} (\vartheta^a)  \ .
\end{equation}

In the following we suppose that fields do not depend on additional coordinates $\vec{x}$ and $\mathbf{z}$. Under the conformal inversion between null infinity and null horizon \eqref{conf dual}, the scalar field transforms according to
\begin{equation}
    \widetilde \Phi (v,r, \vartheta^a ) = \alpha^\delta \Phi(u \to v, r \to r^2_c/r, \vartheta^a) \ .
\end{equation}
This mapping ensures that the multipolar structure is preserved, providing a natural way to relate Aretakis charges to NP charges.

\section{Scalar asymptotic charges in D3-brane backgrounds}\label{GenD3branes}

Let us now discuss how to identify NP and Aretakis charges for D3-branes and their  BPS bound states. Our analysis will rely on generalized CT conformal inversions \cite{Bianchi:2021yqs, Bianchi:2022wku} and on the peculiar properties of D3-branes that involve only the metric and the RR self-dual 4-form, while the complex dilaton and the complex 2-form are inert/inactive.

\subsection{Single D3-brane stack}\label{SingleD3}
The metric for a single stack of $N$ D3-branes extended along the $(t,\vec{x})$ directions reads
\begin{equation}
\label{D3braneMetricXY}
    d\widetilde s^2=H^{-\frac{1}{2}}\left(-dt^2+|d\vec{x}|^2\right)+H^\frac{1}{2}\left(dr^2+r^2d\Omega_5^2\right)\;,\quad H=1+\frac{L^4}{r^4}
\end{equation}
where $L^4=4\pi g_s N\alpha'^2$, $ d\Omega_5^2$ denotes the round metric on the 5-sphere $S^5$ and the self-dual 5-form field strength $F$ is given by
\begin{equation}
\label{D3brane5formXY}
    F = \frac{1}{2}(1+\star)dH^{-1} \wedge dx^0 \wedge dx^1 \wedge dx^2 \wedge dx^3 \ . 
\end{equation}
As already mentioned, the complex dilaton is constant and the two 2-forms vanish. As in the ERN case, it is convenient to work in outgoing $\{u,r,\vec{x},\vartheta^A\}$  or incoming coordinates $\{v,r,\vec{x},\vartheta^A\}$ and write the near-infinity and near-horizon line elements as
\begin{equation}\label{eq: d3 inf}
ds^2_{\mathcal{I}}=-H^{-\frac{1}{2}}du^2-2dudr+H^{\frac{1}{2}}r^2d\Omega_5{}^2 + H^{-\frac{1}{2}}|d \vec{x}|^2 \ ,
\end{equation}
\begin{equation}\label{eq: d3 hor}
ds^2_{\mathcal{H}}=-H^{-\frac{1}{2}}dv^2+2dvdr+H^{\frac{1}{2}}r^2d\Omega_5{}^2+ H^{-\frac{1}{2}}|d \vec{x}|^2  \ .
\end{equation}
Retarded and advanced time are $u_{\Lambda}=t_{\Lambda}-r^*$ and $v_{\Lambda}=t_{\Lambda}+r^*$ where $\Lambda$ is any boost in $D=4$. The tortoise coordinate is defined by $dr^* = H^{1/2}dr$ and turns out to be an elliptic integral
\begin{equation}
r^* = \frac{L}{2} \int_1^{r^2/L^2} \frac{d \xi}{\xi} \sqrt{\xi + \frac{1}{\xi}} 
=\frac{L}{r} \left(r\frac{\sqrt{2 \pi } \Gamma \left(\frac{3}{4}\right)}{\Gamma \left(\frac{1}{4}\right)}-L \, _2F_1\left(-\frac{1}{2},-\frac{1}{4};\frac{3}{4};-\frac{r^4}{L^4}\right)\right)\ ,
\end{equation}
where $_2F_1$ is the hypergeometric function.
Under generalized CT inversions,
\begin{equation}\label{eq: d3 ct}
    u \to v \qquad r \to \frac{L^2}{r} \ ,
\end{equation}
implying that $ r^* \to - r^* $,
the two metrics are dual (related by a Weyl rescaling):
\begin{equation} \label{eq: d3 selfduality}
    ds^2_{\mathcal{I}} \to  \alpha^2 ds^2_{\mathcal{H}} \qquad \alpha= \frac{L}{r} \ .
\end{equation}
This means that the spacetime described by a stack of D3-branes is self-dual in the sense of \cite{Agrawal:2025fsv} and hence it is possible to derive Newman-Penrose (NP) and Aretakis charges and match the ones with the others. As we will see, in order to relate NP and Aretakis charges for fields with spin higher than zero, it necessary to understand how the full solution transforms under CT inversion. By looking at \eqref{D3brane5formXY}, it can be noticed that the 5-form field strength changes sign under such inversion. In particular, this result will be crucial for the spinor analysis in \ref{section: sugra}.

As in the case of extremal RN (ERN) BHs, let us consider a massless inert scalar field $\Phi$, {\it e.g.} the complex dilaton for definiteness\footnote{Note that active scalars and/or other bosonic fields that source a given background mix with each other and with the metric even at the linearized level.}. The near-infinity multipolar expansion of the radiative field is
\begin{equation}
    \Phi(u,r,\vec{x}, \vartheta^a) =\int\frac{d^3p}{(2\pi)^3}e^{i\vec{p} \cdot \vec{x}} \sum_{n=0}\sum_{\ell=0}^{\infty}\sum_{\vec{m} \in M_{\ell}}\frac{1}{r^{n+4}}\varphi^{(n)}_{\ell, \vec{m}, \vec{p}}(u)Y_{\ell, \vec{m}}(\vartheta^a) \ ,\label{eq: expansion d3}
\end{equation}
where the three azymuthal numbers $\vec{m}=(m_1,m_2,m_3) \in M_\ell$ correspond to the three independent projections of the angular momentum along three 2-planes in the 6 transverse directions  and $Y_{\ell, \vec{m}}$ are the spherical harmonics on $S^5$ that satisfy
\begin{equation}
    \nabla_{S^5}^2 Y_{\ell, \vec{m}} = -\ell(\ell+4) Y_{\ell, \vec{m}} \ . 
\end{equation}

Solving the Klein-Gordon equation $\Box_{\mathcal{I}} \Phi = 0$ in outgoing coordinates, suitable near null infinity, one finds
\begin{equation} \label{eq: eom d3}
\begin{gathered}
    2 \partial_u \partial_r \Phi + \frac{1}{\sqrt H r^5} \partial_r (\sqrt H r^5) \partial_u \Phi -\frac{1}{\sqrt H r^5} \Big\{r^3\nabla^2_{
    S^5
    } + \partial_r (r^5 \partial_r) \Big\} \Phi-\sqrt{H}\nabla_{\mathbb{R}^3}^2\Phi= 0 \ .
\end{gathered}
\end{equation}
Expanding the scalar as in  \eqref{eq: expansion d3}, one then obtains per each spherical harmonic, and each plane-wave
\begin{align}  \label{D3modeexp}
    &\sum_{n} \sum_{k=0}^{\infty}\binom{-\frac{1}{2}}{k}L^{4k}\Big[r^{-n-4-4k} L^4(2n+5) \dot{\varphi}^{(n)}_{\ell,\vec{m},\vec{p}}+r^{-n-4k} (2n+3)\dot{\varphi}^{(n)}_{\ell,\vec{m},\vec{p}} \Big] \\ &+ \sum_{n} \Big[ r^{-n-1} [\ell(\ell + 4)-n(n+4) ]+ r^{-n+1} |{\vec{p}}\,|^2 + r^{-n-3} |\vec{p}\,|^2 L^4 \Big] \varphi^{(n)}_{\ell,\vec{m},\vec{p}}=0 \notag \ ,
\end{align}
which implies that each coefficient in the $1/r$ expansion vanishes.  In order to extract an infinite tower of conserved charges, we require that for each power of $r$ the term in the second line of the equation above vanishes,
\begin{equation}
[\ell(\ell + 4)-n(n+4) ]\varphi^{(n)}_{\ell,\vec{m},\vec{p}} + |{\vec{p}}\,|^2 \varphi^{(n+2)}_{\ell,\vec{m},\vec{p}}+  |\vec{p}\,|^2 L^4  \varphi^{(n-2)}_{\ell,\vec{m},\vec{p}} =0 \ .
\end{equation}
For ${\vec{p}} \neq 0$, this imposes a relation among different modes which cannot be satisfied for a generic fluctuation. On the other hand, setting $\vec{p}=0$ and taking $n=\ell$, the condition is automatically satisfied, and as a consequence projecting eq. \eqref{D3modeexp} on such modes  one obtains the Newman–Penrose charges
\begin{equation}
    N_{\ell, \vec{m}} = \sum_{k=0}^{\lfloor \frac{\ell+1}{4} \rfloor}\binom{-\frac{1}{2}}{k} L^{4k} \Big[ L^4 (2\ell-1-8k) \varphi^{(\ell-3-4k)}_{\ell, \vec{m},\vec{p}=0} + (2\ell + 5 - 8k) \varphi^{(\ell+1-4k)}_{\ell, \vec{m},\vec{p}=0}\Big] \ .
\end{equation}
Let us emphasise that the conserved charges require a projection onto modes with vanishing three-momenta. It is  possible to compactify the $\mathbb{R}^3$ directions denoted by $\vec{x}$, in order to discretise the momenta, yet the only solution would remain ${\vec{p}} = 0$. Alternatively, for a fixed momentum one can exploit the asymptotic $SO(1,3)$ Lorentz symmetry to boost the momentum to $\vec{p}=0$ at the cost of boosting the retarded time $u$.

We now repeat the same analysis for the near-horizon expansion, which reads
\begin{equation}
    \widetilde \Phi (v, r, \vec{x}, \vartheta^a) = \int\frac{d^3p}{(2\pi)^3}\sum_{n=0}^{\infty}\sum_{\ell=0}^{\infty} \sum_{\vec{m} \in M_{\ell}} r^n \widetilde \varphi^{(n)}_{\ell, \vec{m},\vec{p}}(v) Y_{\ell, \vec{m}}(\vartheta^a)e^{i \vec{p} \cdot \vec{x}} \ . 
\end{equation}
Again, we focus on the case in which $p_{_I} = 0$, and $n=\ell$. Substituting the expansion above in $\Box_{\mathcal{H}} \widetilde \Phi = 0$ and repeating the same analysis, we find the Aretakis charges
\begin{equation} \label{eq: ar d3}
    A_{\ell, \vec{m}} = \sum_{k=0}^{\lfloor \frac{\ell+1}{4} \rfloor}\binom{-\frac{1}{2}}{k} L^{-4k-2}\Big[ L^4(2\ell + 5 -8k ) \widetilde{\varphi}^{(\ell +1 -4k)}_{\ell, \vec{m}, \vec{p}=0}+(2\ell -1 + 8k) \widetilde{\varphi}^{(\ell - 3 - 4k)}_{\ell, \vec{m}, \vec{p}=0}\Big]\ . 
\end{equation}

The relation between the near-horizon and the near-infinity expansion of the field is
\begin{equation}
    \widetilde \Phi(v,r, \vartheta^a) = \Big(\frac{L}{r}\Big)^4 \Phi (u \to v, r \to L^2/r, \vartheta^a)     \, 
\end{equation}
and hence
\begin{equation}\label{eq: d3 field matching}
    \widetilde \varphi^{(n)}_{\ell,\vec{m},\vec{p}} = \frac{\varphi^{(n)}_{\ell, \vec{m}, \vec{p}}}{L^{2n + 4}} \ ,
\end{equation}
which for $\vec{p}=0$ and $n=\ell$ implies the  matching between NP and Aretakis charges:
\begin{equation}\label{eq:relation NP Aretakis scalar D3}
    A_{\ell, \vec{m}}= \frac{1}{L^{2 \ell + 4}} N_{\ell,\vec{m}} \ .
\end{equation}

\subsubsection{Aretakis instability in D3-branes}
For completeness we summarize the main results found in \cite{Chen:2025sim} concerning the linear Aretakis instability characterizing non-dilatonic extremal black branes. Using symmetry arguments, one can derive the late-time behaviour of transverse derivatives of a generic field at the horizon of D3-branes
\begin{equation} \label{eq: transv deriv}
    \partial^n_{r}\phi|_{r=0} \sim v^{n-\Delta} , \qquad v\to \infty
\end{equation}
where the scaling dimension is given by
\begin{equation}
    \Delta = 2 \left ( 1+ \sqrt{1+ \frac{m^2_{\text{eff}}}{4}}\right) \ . 
\end{equation}
and $m_{\text{eff}}$ is the effective mass that appear in the AdS$_5$ wave equation of the field
\begin{equation}
    (\Box_{\text{AdS$_5$}} - m^2_{\text{eff}}) \phi = 0 \ .
\end{equation}
By looking at \eqref{eq: transv deriv}, one can notice that at least $\lfloor \Delta \rfloor$ or $\lfloor \Delta  \rfloor+1$ derivatives are necessary to see respectively non-decay or power-law blow up in the advanced time $v$. This signals the presence of Aretakis instability.

\subsection{D3-D3$'$ background}\label{sec:D3-D3}
\begin{table}[]
\centering
             \begin{tabular}{c|cccccccccc}
                   & $t$&$y_1$&$y_2$&$y_3$&$y_4$&$x$&$z_1$&$z_2$&$z_3$&$z_4$ \\\hline
                   D3& $\fatdash$&$\boldsymbol{\cdot}$&$\boldsymbol{\cdot}$&$\boldsymbol{\cdot}$&$\boldsymbol{\cdot}$&$\fatdash$&$\fatdash$&$\fatdash$&$\boldsymbol{\cdot}$&$\boldsymbol{\cdot}$\\
                   D3$'$ & $\fatdash$&$\boldsymbol{\cdot}$&$\boldsymbol{\cdot}$&$\boldsymbol{\cdot}$&$\boldsymbol{\cdot}$&$\fatdash$&$\boldsymbol{\cdot}$&$\boldsymbol{\cdot}$&$\fatdash$&$\fatdash$
             \end{tabular}
                 \caption{Table illustrating the positions of the D3 and D3$'$ stacks. }
                 \label{tab: TwoD3stacks}
\end{table}
Consider two stacks of D3-branes. The first stack is along $\{t,x;z_1,z_2\}$, the second is along $\{t,x;z_3,z_4\}$, as illustrated in Table \ref{tab: TwoD3stacks}. Assuming for simplicity that the number of D3-branes in each stack is the same $N_1=N_2=N$, the metric depends on a single  harmonic function
\begin{equation}
    H= 1 + {L^2\over r^2} \ , 
\end{equation}
where $L^2 = 4\pi g_s N \alpha'^2/(2\pi R)^2$ and $r^2 = \sum_i y_i^2$ is the radial coordinate in the four transverse directions. The corresponding solution is
\begin{equation}
    ds^2=H^{-1}(-dt^2+dx^2)+H\left(dr^2+r^2d\Omega^2_3\right)+d\textbf{z}^2 \ . 
\end{equation}
where $d\Omega^2_3$ denotes  the line element on a unit $S^3$. In outgoing coordinates, suitable for the near-infinity analysis, the D3-D3$'$-brane metric reads
\begin{equation}\label{eq:D3-D3' near infinity metric}
    ds^2_{\mathcal{I}} = - H^{-1} du^2 - 2 du dr + r^2 H d\Omega^2_3 + H^{-1}dx^2 + d\mathbf{z}^2 \ , 
\end{equation}
where $u=t-r^*$, with 
\begin{equation}
    r^* = r- {L^2\over r}
\end{equation}
is the tortoise coordinate, chosen so that $r^*=0$ for $r=L$ (light-ring). 

The Klein-Gordon equation for a massless scalar field, such as the Type-IIB complex dilaton, or more generally a complex probe scalar, in the near-infinity background geometry takes the form
\begin{equation}
\left[
-\left(\frac{H'}{H}+\frac{3}{r}\right)\partial_u
-2\,\partial_u\partial_r
+\frac{3}{rH}\partial_r
+\frac{1}{H}\partial_r^2
+\frac{1}{r^2 H}\nabla_{S^3}^2
+H\,\partial_x^2
+\partial_{\mathbf{z}}^2
\right]\Phi = 0 \ .
\end{equation}

The large $r$ expansion of the field is the following:
\begin{equation}
    \Phi(u,r,x, \vartheta, \phi, \psi,\mathbf{z}) = \int\frac{dp}{2\pi}\sum_k\sum_{n=0} \sum_{\ell=0}^{}\sum_{m,m'=-\ell/2}^{\ell/2}\frac{\varphi^{(n)}_{\ell, p, \vec{k}}(u)}{r^{n+2}}D_{\ell, m,m'}(\vartheta, \phi, \psi)e^{-ipx}e^{-ik_iz^i} \ ,
\end{equation}
where $p$ and $\vec{k}$ are respectively the Kaluza-Klein momentum along the $x$-direction and the $\mathbf{z}$-directions. The $D_{\ell,m,m'}$ are the spherical harmonics on $S^3$, which are equivalent to Wigner's functions:
\begin{equation}
    D^\ell_{m,m'}(\vartheta, \phi, \psi)=  d^\ell_{m,m'}(\vartheta) e^{im\phi+im'\psi} \ ,
\end{equation}
satisfying
\begin{equation}
    \nabla_{S^3}^2D^{\ell}_{m,m'}=-\ell(\ell+2)D^{\ell}_{m,m'} \ .
\end{equation}

In order to derive conserved charges, one must choose the momenta along the common longitudinal direction ${x}$  to vanish. Similarly, the momenta $\vec{k}$ along the $\textbf{z}$ directions are also required to vanish in order to obtain a conservation equation.
Substituting the expansion into the equation of motion we arrive at
\begin{align*}
        \frac{1}{r^2 + L^2} \Big\{& \sum_n\sum_\ell \Big[ -\ell(\ell+2) + n(n+2) \Big] r^{-n-2} \varphi^{(n)}_\ell D_\ell + L^2 \sum_n\sum_\ell (3+2n) r^{-n-3} \dot \varphi^{(n)}_\ell D_\ell \\
        & + \sum_n \sum_\ell (1+2n) r^{-n-1} \dot \varphi^{(n)}_\ell D_\ell\Big\} = 0  \ .
\end{align*} 
We notice that for $n=\ell$ the first term vanishes, and repeating the same arguments as in the previous cases
we obtain the NP conserved charges
\begin{equation}
    N_{\ell,m,m'} = L^2(2\ell+1) \varphi^{(\ell-1)}_{\ell,m,m',p=0,\vec{k}=0} + (2\ell+3)\varphi^{(\ell+1)}_{\ell,m,m',p=0,\vec{k}=0} \ .  
\end{equation} 
The  $D3-D3'$ spacetime is self-dual under conformal inversion $u=v$ and $r=L^2/r$ in the sense that the six-dimensional metric obtained changes by a Weyl rescaling
 \begin{equation}
     ds^2_{\mathcal{I}} - d \mathbf{z}^2 \to \alpha^2 (ds^2_{\mathcal{H}} - d \mathbf{z}^2 )\;,
 \end{equation}
where $\alpha = L/r$ and
\begin{equation}
    ds^2_{\mathcal{H}} = -H^{-1} dv^2 + 2dvdr + r^2 H d\Omega^2 + H^{-1}dx^2 + d\mathbf{z}^2
\end{equation}
is the near-horizon line element.
Now we use the following expansion for the scalar field (imposing $p=0$ and $k_i=0$):
\begin{equation}
    \widetilde{\Phi}(v,r,\vartheta, \phi,\psi) = \sum_{n=0}^{\infty} \sum_{\ell=0}^{\infty} \sum_{m, m'} r^n \widetilde{\varphi}^{(n)}_{\ell, p, \vec{k}}(v)D_\ell(\vartheta, \phi, \psi) \ .
\end{equation}
Repeating the same computation as before, at the end we find the  Aretakis charges
\begin{equation}\label{aretakis}
    A_{\ell,m,m'} =  L^2 (2\ell+3) \widetilde{\varphi}^{(\ell+1)}_{\ell,m,m',p=0, \vec{k}=0} + (2\ell+1) \widetilde{\varphi}^{(\ell-1)}_{\ell,m,m',p=0, \vec{k}=0}  \ .
\end{equation}

Under conformal transformation, scalar fields transform as
\begin{equation}
   \widetilde{\Phi} (v,r,\vartheta, \phi, \psi) = \alpha^{2} \Phi(u \to v, r\to L^2/r, \vartheta, \phi, \psi) \qquad \alpha = L/r \ ,
\end{equation}
so that the near-horizon and near-infinity coefficients are related by 
\begin{equation}\label{match_1}
    \widetilde{\varphi}^{(n)}_\ell = \frac{1}{L^{2n+2}}\varphi^{(n)}_\ell \ ,
\end{equation}
and substituting Eq. \eqref{match_1} into Eq. \eqref{aretakis},  one finds the relation between NP charges and Aretakis charges
\begin{equation}
    A_{\ell,m,m'} = \frac{1}{L^{2\ell+2}}N_{\ell,m,m'} \ .
\end{equation}

\subsection{Four D3-brane stacks and Extreme BHs in STU supergravity}\label{FourD3STU}
The “large” extremal black hole solution in STU supergravity may be described as
the BPS intersection of four stacks of  D3-branes as in Table \ref{tab: 4stackD3}, with each stack carrying charge $q_i$. 

\begin{table}[]
\centering
             \begin{tabular}{c|cccccccccc}
                   & $t$&$y_1$&$y_2$&$y_3$&$z_1$&$z_2$&$z_3$&$z_4$&$z_5$&$z_6$ \\\hline
                   D3& $\fatdash$&$\boldsymbol{\cdot}$&$\boldsymbol{\cdot}$&$\boldsymbol{\cdot}$&$\fatdash$&$\boldsymbol{\cdot}$&$\fatdash$&$\boldsymbol{\cdot}$&$\fatdash$&$\boldsymbol{\cdot}$\\
                   D3$'$ &$\fatdash$&$\boldsymbol{\cdot}$&$\boldsymbol{\cdot}$&$\boldsymbol{\cdot}$&$\boldsymbol{\cdot}$& $\fatdash$ &$\boldsymbol{\cdot}$& $\fatdash$ & $\fatdash$ &$\boldsymbol{\cdot}$\\
                   D3$''$ & $\fatdash$&$\boldsymbol{\cdot}$&$\boldsymbol{\cdot}$&$\boldsymbol{\cdot}$&$\fatdash$ &$\boldsymbol{\cdot}$&$\boldsymbol{\cdot}$& $\fatdash$ &$\boldsymbol{\cdot}$& $\fatdash$ \\
                   D3$'''$ &$\fatdash$&$\boldsymbol{\cdot}$&$\boldsymbol{\cdot}$&$\boldsymbol{\cdot}$&$\boldsymbol{\cdot}$& $\fatdash$ & $\fatdash$ &$\boldsymbol{\cdot}$&$\boldsymbol{\cdot}$& $\fatdash$ \\
             \end{tabular}
                 \caption{The positions of the four stacks of D3-branes.}\label{tab: 4stackD3}
\end{table}

The full metric is
\begin{equation}\label{eq: stu metric}
ds^2 = -H^{-1/2}dt^2 + H^{1/2}(dr^2 + r^2 d\Omega_{S^2}^2) + ds_{T_6}^2 \,
\end{equation}
where $r^2 = \sum_iy_i^2$, the warp factor is
\begin{equation}
    H = \prod_{i=1}^4H_i \ \ \ \ \ \ \ \ \ \ H_i = 1 + \frac{q_i}{r} \ ,
\end{equation}
and $ds^2_{T_6}$ is the metric of the six-dimensional torus spanned by $z$-coordinates,
\begin{align}
    ds^2_{T_6} &= \Big(\frac{H_2H_4}{H_1H_3}\Big)^{1/2} dz_1^2 + \Big(\frac{H_1H_3}{H_2H_4}\Big)^{1/2} dz_2^2 + \Big(\frac{H_2H_3}{H_1H_4}\Big)^{1/2}dz^2_3 \notag \\
    &+ \Big( \frac{H_1 H_4}{H_2 H_3} \Big)^{1/2} dz^2_4 + \Big( \frac{H_3 H_4}{H_1 H_2} \Big)^{1/2} dz^2_5 +\Big( \frac{H_1 H_2}{H_3 H_4} \Big)^{1/2} dz^2_6   \ . 
\end{align}

We can define outgoing and ingoing coordinates $u=t-r^*$ and $v=t+r^*$ where $dr^*/dr = H^{1/2}$. The near-infinity metric is
\begin{equation}\label{eq: stu metric inf}
    ds^2_{\mathcal{I}} = - H^{-1/2} du^2 - 2dudr + H^{1/2} r^2 d\Omega_{S^2}^2 + ds^2_{T_6} \ ,
\end{equation}
while the near-horizon metric is
\begin{equation}\label{eq: stu metric hor}
    ds^2_{\mathcal{H}} = - H^{-1/2} dv^2 + 2dvdr + H^{1/2} r^2 d\Omega_{S^2}^2 + ds^2_{T_6} \ .
\end{equation}

Let us consider the CT inversion
\begin{equation} \label{eq: stu map}
   u \to  v \qquad  r \to \frac{Q^2}{r} \qquad Q^4=q_1q_2q_3q_4 \ .
\end{equation}
The four-dimensional part of the metric \eqref{eq: stu metric inf} transforms as
\begin{equation}\label{eq: stu conf}
    ds^2_{\mathcal{I}} - ds^2_{T_6} \to \alpha^2 (d\widetilde s^2_{\mathcal{H}} -ds^2_{T_6})\ \ \ \ \ \ \ \ \alpha = Q/r \ ,
\end{equation}
where
\begin{equation}
    d\widetilde s^2_{\mathcal{H}} = - \widetilde H^{-1/2} dv^2 + 2 dv dr + \widetilde H^{1/2} r^2 d \Omega_{S^2}^2 + ds^2_{T_6}
\end{equation}
and
\begin{equation}
    \widetilde H = \prod_{i=1}^4\widetilde H_i \qquad \widetilde H_i = 1 + \frac{\widetilde q_i}{r} \qquad  \widetilde q_i = {Q^2\over q_i} \ .
\end{equation}

\subsubsection{Pairwise equal charges}
Since we need to redefine charges to obtain the conformal relation between the two metrics, the inversion is not a conformal symmetry. However, there exist cases in which the map \eqref{eq: stu map} is a conformal symmetry, namely when the four electric charges $\widetilde{q}_i$ are a permutation of the $q_i$. This happens for instance for pairwise equal charges, for example $q_1=q_3=q$ and $q_2=q_4=\hat q$.

In this case the line elements become
\begin{equation}
    ds^2_{\mathcal{I}} = - H^{-1} du^2 - 2dudr+Hr^2d\Omega_{S^2}^2 + ds^2_{T_6}
\end{equation}
\begin{equation}
    ds^2_{\mathcal{H}} = - H^{-1} dv^2 + 2dvdr+Hr^2d\Omega_{S^2}^2 + ds^2_{T_6}
\end{equation}
where
\begin{equation}
    H = \Big(1+\frac{q}{r}\Big)\Big(1+\frac{\hat q}{r}\Big)
\end{equation}
and the compactified metric is
\begin{equation}
    ds^2_{T_6} = \Big( \frac{1+ \hat q/r}{1+ q/r} \Big) dz_1^2 + \Big(\frac{1 + q/r }{1 + \hat{q}/r} \Big) dz^2_2 + dz^2_3 + dz^2_4 + dz^2_5 + dz^2_6 \ .
\end{equation}
Now the map \eqref{eq: stu map} is a conformal symmetry between near-horizon and near-infinity spacetimes, with a factor
\begin{equation}
    \alpha = \frac{\sqrt{q \hat q}}{r} 
\end{equation}

Let us consider a massless scalar field $\Phi$ with zero momentum along the $z$-directions. The large $r$ and small $r$ expansions are the following:
\begin{equation} \label{eq: stu inf exp}
    \Phi(u,r,\vartheta,\phi) = \sum_{n=0} \sum_{\ell=0}\sum_{m=-\ell}^{+\ell} \frac{1} {r^{n+1}}\varphi^{(n)}_{\ell, m}(u) Y_{\ell, m}(\vartheta, \phi)  \ ,
\end{equation}
\begin{equation} \label{eq: stu hor exp}
    \widetilde \Phi(v,r,\vartheta,\phi) = \sum_{n=0} \sum_{\ell=0}\sum_{m=-\ell}^{\ell}r^n \widetilde \varphi^{(n)}_{\ell,m}(v) Y_{\ell,m}(\vartheta, \phi) \ . 
\end{equation}
As usual $Y_{\ell, m}$ are the spherical harmonics on $S^2$. Substituting \eqref{eq: stu inf exp} in $\Box_\mathcal{I} \Phi =0 $ we find
\begin{align}
    \sum_{n, \ell, m}&\Big[ -\ell(\ell+1) + n(n+1) \Big] r^{-n-1} \varphi^{(n)}_{\ell.m} Y_{\ell.m} +2q \hat q\sum_{n, \ell, m} (n+1) r^{-n-2} \varphi^{(n)}_{\ell, m} Y_{\ell, m}  \notag \\
    &+ \sum_{n, \ell, m} (q + \hat q)(2n+1) r^{-n-1} \varphi^{(n)}_{\ell, m} Y_{\ell.m} + 2\sum_{n, \ell, m} n \varphi^{(n)}_{\ell,m} Y_{\ell,m}  =0 \ , 
\end{align}
from which we can read the NP charges
\begin{equation}
 N_{\ell,m} =  2 q \hat q \ell \varphi^{(\ell-1)}_{\ell, m} + (2\ell+1)(q + \hat q) \varphi^{(\ell)}_{\ell,m} + 2(\ell+1) \varphi^{(\ell+1)}_{\ell,m} \ .
\end{equation}

Analogously, substituting \eqref{eq: stu hor exp} in $\Box_{\mathcal{H}} \Phi = 0$, we find the Aretakis charges
\begin{equation}
    A_{\ell, m} = 2q \hat q(\ell+1) \widetilde \varphi^{(\ell+1)}_{\ell,m} + (2\ell+1)(q + \hat q) \widetilde \varphi^{(\ell)}_{\ell,m} + 2\ell \widetilde \varphi ^{(\ell-1)}_{\ell,m}      .
\end{equation}

Under the CT inversion the scalar field transforms as
\begin{equation}
    \widetilde \Phi (v,r,\vartheta,\phi) = \frac{\sqrt{q \hat q}}{r} \Phi(u \to v, r \to q \hat q / r, \vartheta, \phi ) \ ,
\end{equation}
and hence
\begin{equation}
    \widetilde \varphi^{(n)}_\ell = \frac{1}{(q \hat q)^{\frac{2n+1}{2}}}\varphi^{(n)}_\ell \ . 
\end{equation}
Finally, the matching between charges is 
\begin{equation}
    A_{\ell,m} = \frac{1}{(q \hat q)^{\frac{2 \ell+1}{2}}} N_{\ell,m} \ .
\end{equation}

\subsubsection{All equal charges}
In the particular case in which $q=\hat q$, one finds 
\begin{equation}
    N_{\ell,m}= 2q^2\ell\varphi^{(\ell-1)}_{\ell,m} + 2q(2\ell+1) \varphi^{(\ell)}_{\ell,m} + 2(\ell+1)  \varphi^{(\ell+1)}_{\ell,m} \ ,
\end{equation}
\begin{equation}
    A_{\ell,m} = 2q^2(\ell+1) \widetilde \varphi^{(\ell+1)}_{\ell,m} +2q(2\ell+1) \widetilde \varphi^{(\ell)}_{\ell,m} + 2\ell \widetilde \varphi^{(\ell-1)}_{\ell, m} \ , 
\end{equation}
and the matching condition becomes
\begin{equation}
    A_{\ell,m}= \frac{1}{q^{2\ell+1}} N_{\ell,m} \ . 
\end{equation}
These are exactly the results that we found in the extremal RN case, and indeed this last equation is identical to eq. \eqref{eq: ANPERN} observing that $q=M$ in the extremal case.

\section{Spinor and tensor asymptotic charges in Type-IIB supergravity}\label{section: sugra}

In this Section we first introduce the fundamental features of Type-IIB supergravity, namely the BPS projection, which halves the number of independent Killing spinors, and the supersymmetric transformations relating the bosonic and fermionic fields of the theory. 
We then analyse the NP and Aretakis charges in this supersymmetric setting. Starting form scalar charges associated with the dilaton fluctuations, it is possible, by means of supersymmetry transformations, to construct spinorial charges associated with the dilatino fluctuations. A similar procedure can be employed to construct asymptotic charges for the higher components of the supergravity multiplet, up to the ones associated to the graviton. 
This analysis is motivated by an interesting relation between the structure of the on-shell superfield in $\mathcal{N} = (2,0) $ supersymmetry and the form of what would correspond to NP and Aretakis conserved charges. 

To make this proposal more concrete, we explicitly compute the near-infinity and near-horizon charges associated with the dilatino and verify the matching between them via Couch-Torrence inversion. 

\subsection{Supersymmetry and BPS conditions}\label{SUSY}

Consider ten-dimensional Type-IIB supergravity, the low-energy effective description of Type-IIB superstring theory. Its scalar sector parametrizes the coset manifold $SL(2,\mathbb{R)}/SO(2) \simeq SU(1,1)/U(1)$, where $SL(2, \mathbb{R})\simeq SU(1,1)$ is a non-compact global symmetry of the low energy theory \cite{Schwarz:1983qr, Bergshoeff:2005ac}. \footnote{Including higher-derivative perturbative and non-perturbative terms $SL(2, \mathbb{R})$ is expected to be broken to $SL(2, \mathbb{Z})$. }

The bosonic field content includes the zehnbein $E^{\hat{A}}_{M}$, two scalars $V_{\pm}^{\alpha}$ that parametrize the coset $SU(1,1)/U(1)$, two 2-forms $B^{\alpha}_{MN}$ and a self-dual 4-form $A^{+}_{M N P Q}$ while the fermion field content includes a complex left-handed gravitino $\Psi_{M}$ and a complex right-handed spinor (the dilatino) $\Lambda$,
\begin{equation}
    \Gamma_{11}\Psi_{M} = - \Psi_{M} \qquad \Gamma_{11}\Lambda = +\Lambda\;,
\end{equation}
where $\Gamma_{11}$ is the chirality operator (equivalent of $\gamma_5$ in $D=10$) and $M=0,\dots,{9}$ . Throughout, we work in the Dirac basis where spinors in ten dimensions are 32-component objects. Appendix \ref{appendixA} contains an in-depth summary of notation and conventions used throughout this section, including spinor notation, and basis of gamma-matrices.

The supersymmetry transformations are
\begin{align}
    \delta_{\epsilon}{V}^\alpha_{+} &= {V}^\alpha_{-} \bar\epsilon^*\Lambda \ , \\
    \delta_{\epsilon}{V}^\alpha_{-} &= {V}^\alpha_{+} \bar\epsilon\Lambda^* \ ,\\
        \delta_{\epsilon}\Lambda &= i P_M \Gamma^M \epsilon \ , \\ 
            \delta_{\epsilon} B^\alpha_{MN} &= V^\alpha_{-}\bar\epsilon\Gamma_{MN} \Lambda + V^\alpha_{+}\bar\epsilon^*\Gamma_{MN}\Lambda^* + 4i V^\alpha_{-}\bar\epsilon\Gamma_{[M}\Psi_{N]} + 4i V^\alpha_{+}\bar\epsilon^*\Gamma_{[M} \Psi_{N]}^* \ , \\
            \delta_{\epsilon} \Psi_{M} &= D_{M} \epsilon + \frac{i}{960}\slashed{F}\Gamma_{M}\epsilon \ , \\
            \delta_{\epsilon}{E^{\hat{A}}_M} &= i \bar\epsilon\Gamma^{\hat{A}}\psi_M + i \bar\epsilon^*\Gamma^{\hat{A}}\psi_M^* \ , \\
            \delta_{\epsilon} A_{MNRS}^{+} &= \bar\epsilon\Gamma_{[MNR}\Psi_{S]} -\bar\epsilon^*\Gamma_{[MNR}\Psi_{S]}^* - {3\over 8}i \varepsilon_{\alpha\beta} B^\alpha_{[MN}\delta{B}^\beta_{RS]} \ . 
\end{align}
where $P_M = \varepsilon_{\alpha\beta} V^\alpha_{+}D_M V^\beta_{+}$ encodes the kinetic contribution of the complex dilaton $(\Phi, \chi)$ and $F$ is the five-form field-strength associated to $A^+$ (see \eqref{D3brane5formXY}). We denote with the asterix $^*$ the complex (Majorana) conjugate of a spinor.

\begin{table}[]
    \centering
    \begin{tabular}{l|r}
        Fields & $U(1)_R$ charge  \\
        \hline
        $A^+_{MNPQ}$ & $0$ \\
        $E^{\hat{A}}_M$ & 0 \\
        $\epsilon$ & $1/2$ \\
        $\psi_M$ & $1/2$ \\
        $V^{\alpha}_{\pm}$ & $\pm 1$ \\
        $B_{MN}^{\alpha}$ & $0$ \\
        $\Lambda$ & $3/2$ \\
        $P_M \sim \Phi, \chi$& $2$ \\
    \end{tabular}
    \caption{$U(1)_R$ charges of various fields in Type-IIB supergravity.}
    \label{tab:r charges}
\end{table}

As seen before, Type-IIB supergravity admits a 1/2 BPS solution representing a stack of $N$ D3-branes
with metric \eqref{D3braneMetricXY} and self-dual 5-form \eqref{D3brane5formXY}. The complex dilaton ($\Phi$, $\chi$) is constant, and the gravitino, dilatino and 2-forms vanish. In order to identify the residual global supersymmetry of the background, the only relevant supersymmetry transformation is the gravitino's:
\begin{equation} \label{eq: susy transf psi}
    \delta_{\epsilon} \Psi_{M} = D_{M} \epsilon + \frac{i}{960}\slashed{F}\Gamma_{M}\epsilon  \ .
\end{equation}

Here $\epsilon$ is a complex left-handed spinor satisfying $\Gamma_{11}\epsilon = - \epsilon $ and $\slashed{F} = F_{N_1\dots N_5}\Gamma^{N_1\dots N_5}$, where $F_{M_1 \dots M_5}$  
are the components of the self-dual 5-form $F$ \eqref{D3brane5formXY}. The covariant derivative reads
\begin{equation}
D_ {M} = \partial_{M} + \frac{1}{4} \Omega_{M}^{\hat N \hat P}\Gamma_{\hat N \hat P}    \ .
\end{equation} 
The spin connection can be expressed in terms of the coordinate-covariant derivative of the zehn-bein as:
\begin{equation}
    E^{\hat Q}_N \nabla_M E^N_{\hat P} = \Omega^{\hat Q}_{M \hat P} \ .
\end{equation}
Preservation of supersymmetry requires the gravitino variation \eqref{eq: susy transf psi} to vanish. Not without effort one can show that the Killing spinor is \cite{Ortin:2015hya}
\begin{equation} \label{eq: bps conditions}
    \epsilon_{_{BPS}} = H^{-1/8} \epsilon_0 \qquad \Big(1 \mp \im\Gamma^{0123} \Big) \epsilon_0 = 0  \ ,
\end{equation}
where $\epsilon_0$ is a constant spinor and the choice of sign in the second expression  corresponds to a choice of D3-brane or $\bar{\mathrm{D3}}$-brane. Henceforth we will drop the subscript BPS of the Killing spinor. The projection reduces the number of independent components of the constant spinor. Physically, this corresponds to the fact that the D3-brane preserves only half of the original Type-IIB supersymmetries. In other words, out of the 32 supercharges of the theory, only 16 remain unbroken in the presence of a flat D3-brane. 

At null infinity one has flat spacetime and the 16 broken supercharges get restored. Near the horizon one gets  AdS$_5$$\times S^5$ whereby 16 new superconformal charges appear that generate the superalgebra/group $PSU(2,2|4)$ together with 15+15 generators of $SO(2,4)\sim SU(2,2)$ (conformal group to be identified with the isometry of AdS$_5$) and of $SO(6)\sim SU(4)$ (R-symmetry group to be identified with the isometry of $S^5$). The common subset of generators of the two should consist of 16 supercharges and 15+6+4 bosonic generators $SO(6)\sim SU(4)$ and $ISO(1,3)\sim SO(1,3) \rtimes T_4 \subset SO(1,4)$, that may result from an Inönü-Wigner contraction \cite{Ruzziconi:2026bix}.

Computing explicitly \eqref{eq: susy transf psi} in outgoing coordinates \eqref{eq: d3 inf}  
\begin{equation}
    (1 -\widetilde{\gamma}_5 ) \epsilon = 0 \ ,
\end{equation}
where $\Tilde{\gamma}_5$ is defined in appendix \ref{appendixA}.
Applying CT inversion \eqref{eq: d3 ct} to the outgoing line element yields the ingoing line element \eqref{eq: d3 selfduality}. Solving the Killing spinor equation in these coordinates leads to
\begin{equation}
    (1 + \widetilde{\gamma}_5 ) \epsilon = 0 \ . 
\end{equation}
This shows that CT conformal inversion flips the eingevalue of $\widetilde{\gamma}_5$ acting on the Killing spinor.  Since the sign of this projection determines the orientation of the five-form flux and therefore the brane charge, this transformation maps a D3-brane configuration into an anti-D3-brane configuration. The two families of solutions can therefore be labelled by $\eta = \pm 1$ where $\eta = +1$ corresponds to the D3-brane solution while $\eta = -1$ to the anti-D3-brane one.

Working in coordinates $X^M = (x^{\tilde \mu}, \vartheta^a)$ where $x^{\tilde \mu} = (u, \vec{x}, r)$ or $x^{\tilde \mu} = (v, \vec{x}, r)$, one can in principle decompose the Killing spinor as \cite{Kim:1985ez}
\begin{equation} \label{eq: killing spinor deco}
    \epsilon = \sum_{\eta} \varepsilon_{\eta}(x^{\tilde \mu}) \kappa_{\eta} (\vartheta^a) \ ,
\end{equation}
where $\kappa_{\eta}$ is the Killing spinor on $S^5$ satisfying \eqref{eq: k s5}.
The Killing spinor equations then select different sectors depending on the choice of coordinates. In outgoing coordinates only the component with $\eta = +1$ survives, while in ingoing coordinates only the component with $\eta = -1$ is allowed.

\subsection{NP and Aretakis superfield}\label{NPASuperfield}

In Type-IIB supergravity, expanding around the D3-brane solution, one can prove that in addition to the scalar charges one can find spinorial and tensorial charges that are preserved at the linearized and are related to the scalar charges by unbroken supersymmetry transformations.

Recall that ${\cal N}=(2,0)$ SUSY admits an on-shell superfield that is chiral and may be used to build 1/2 BPS terms in the effective action like the famous $R^4$ term and SUSY related terms. It can be written as
\begin{equation}
\begin{split}
    \widehat\Phi(X,\theta) =& \Phi(X) + \bar{\theta}^*\Lambda(X)  + {1\over 3!} \bar{\theta}^*\Gamma^{LMN}\theta [G_{LMN}(X)   + \bar{\theta}^*\Gamma_L D_M  \Psi_N(X) \\& + \bar{\theta}^*\Gamma^{KPR}\theta (g_{LK} W_{MNPR}(X)  + D_L F_{MNKPR}(X) )] + ...
\end{split}
\end{equation}
where $
  \bar{\theta}^*\Gamma^{MNP}\theta
$ can be written in the Weyl basis as $ \theta^\alpha \hat{\Gamma}^{MNP}_{\alpha\beta}\theta^\beta
$ which is the only Lorentz-invariant combination of two anticommuting spinors $\theta^\alpha$ in the ${\bf 16}$ of $SO(1,9)$. 
Note that all the gauge fields (including the gravitino!) appear  only through their field strengths. 
Moreover the graviton and 4-form appear with two-derivatives (at least) at the same order in the expansion. Last but not least with the standard $U(1)_R$ charge assignments (see Tab. \ref{tab:r charges}), all terms have charge $+2$ if one assigns charge +1/2 to $\theta$'s. 

Denoting by $N^\Phi$ the scalar charges, by $N^\Lambda$ the dilatino charges, by $N^{B}$ the complex antisymmetric tensor charges, by $N^\Psi$ the gravitino charges, by $N^{A}$ the (self-dual) 4-form charges and by $N^H$ the spin-2 graviton charges one expects to find (schematically): 

\begin{align}
    N^{\Phi} &\sim  \Phi \ ,  \\
    N^{\Lambda} &\sim  \bar\epsilon^*\Lambda \ ,   \\
    N^{B} &\sim  \bar\epsilon^*\Gamma_{MNP}\epsilon G^{MNP} \ ,\\
    N^{\Psi} &\sim  \bar\epsilon^* \Gamma^{MNP}\epsilon \bar\epsilon^* \Gamma_{[M} D_N \Psi_{P]}  \ , \\
    N^{A} &\sim  
\bar\epsilon^* \Gamma^{MNP}\epsilon 
\bar\epsilon^* \Gamma^{QRS}\epsilon D_{[M} F_{NPQRS]}  \ ,  \\
N^{H} &\sim  
g_{PQ}\bar\epsilon^* \Gamma^{MNP}\epsilon 
\bar\epsilon^* \Gamma^{QRS}\epsilon W_{MNRS} \ , 
\end{align}
where $W_{MNRS}$ is the Weyl tensor and $ G_{MNP}= V_\alpha^+ D_{[M} B^\alpha_{NP]}$.
Notice that the NP charges associated to fields transforming under local symmetries are gauge invariant since they only involve their field-strengths.

Notice the tantalizing similarity between the various conserved (and gauge invariant) charges and the components of the on-shell superfield, after the replacement $\theta^\alpha \rightarrow \epsilon^\alpha$. This is \emph{necessary} since $\theta$ are anticommuting, while $\epsilon$ is a commuting spinor. Hence one can define a super-field progenitor of the NP (and Aretakis) charges 
\begin{equation}
  \widehat{N}(X, \epsilon) = \widehat{\Phi}(X, \theta=\epsilon) \ . 
\end{equation}
The actual NP/Aretakis charges require decomposing $\widehat{\Phi}(X,\theta)$ at each order in $\theta\sim \epsilon/\bar\epsilon^*$ in spherical harmonics and combining the terms that can mix. 
Schematically (see KK reduction on $S^5$ in \ref{Appendix: KK modes}) one finds \cite{Gunaydin:1984fk, Kim:1985ez}: 
\begin{itemize}
\item $\Phi$: One set of modes associated to scalar spherical harmonics $\Delta=4+\ell_1$
\item $\Lambda$: Two sets of spinor modes associated to spinor spherical harmonics $\Delta={7\over 2} +\ell_4$ and $\Delta=-{3\over 2} +\ell_4$
\item $B_{\mu\nu}$: Two sets of anti-symmetric tensor modes associated to scalar spherical harmonics $\Delta=2+\ell_1$ (with $\ell_1\ge 1$) and $\Delta=6+\ell_1$ 
\item $B_{\mu a}$: One set of vector modes associated to vector spherical harmonics $\Delta = 5+\ell_5$
\item $B_{ab}$: One set of scalar modes associated to anti-symmetric tensor spherical harmonics $\Delta = 3+\ell_{10}$
\item $\Psi_\mu$: Two set of spin-vector modes associated to spinor spherical harmonics with $\Delta={7\over 2} +\ell_4$ and $\Delta={13\over 2} +\ell_4$
\item $\Psi_a$: Four set of spinor modes associated to spinor-vector spherical harmonics $\Delta={7\over 2} +\ell_{20}$, $\Delta=-{1\over 2} +\ell_{20}$ and two to the internal $\Gamma$-trace  with $\Delta={5\over 2}+\ell_4$ and $\Delta={11\over 2}+\ell_4$
\item $H_{\mu\nu}$: One set of symmetric traceless tensor modes associated to scalar spherical harmonics $\Delta=\ell_1+4$
\item $H_{(ab)}$: One set of scalar md associated to tensor spherical harmonics $\Delta=\ell_{14}+6$
\item $H_{a\mu}\pm A_{abc \mu}$: Two sets of vector modes associated to vector spherical harmonics $\Delta=\ell_{5}+3$ and $\Delta=\ell_{5}+6$ 
\item $H^a{}_a\pm A_{abc d}$ : Two sets of scalar modes associated to scalar harmonics with $\Delta=\ell_1$ ($\ell_1\ge 2$) and $\Delta=\ell_1+8$
\item $A_{ab\mu\nu}$ : One set of charges associated to anti-symmetric tensor harmonics with $\Delta=\ell_{10}+3$ 
\end{itemize}

In order to substantiate our claim, let us  consider a small fluctuation of the dilaton field, denoted by $\delta \Phi$. In the asymptotic region near spatial infinity, the fluctuation can be expanded in multipolar modes as
\begin{equation}
 \delta \Phi(u,r,\vartheta^a) = \sum_{n=0} \sum_{\ell=0}^{\infty}\sum_{\vec{m} \in M_{\ell}} \frac{1}{r^{n+4}} \delta \varphi^{(n)}_{\ell, \vec{m}}(u) Y_{\ell, \vec{m}}(\vartheta^a)
\end{equation}
where we set to zero the momentum along longitudinal spatial directions. The dilaton fluctuation induces a non-trivial dilatino via supersymmetry transformation:
\begin{equation}
\delta_{\epsilon} \Lambda = \Gamma^M \partial_M \delta \Phi \epsilon\;.
\end{equation}
In the next subsection, we will consider the equations of motion for the dilatino in the asymptotic regions and verify their consistency with this expression.

\subsection{Asymptotic charges from the dilatino}

In this subsection we explicitly identify the dilatino charges. The reader will find a summary of the notations used throughout here in appendix \ref{appendixA}. The linearized equation for the fluctuation mode of the complex dilatino $\Lambda $ reads \cite{Schwarz:1983qr}
\begin{equation}\label{eq: dilatino eq}
    \slashed{D}\Lambda-\frac{i}{480}\slashed{F_5}\Lambda =0\;.
\end{equation}

We begin by working in the outgoing coordinates \eqref{eq: d3 inf}. For later use let us quote 
\begin{equation}
    \frac{1}{480}\slashed{F_5}=\frac{1}{8}\left(\frac{H'}{H}\Gamma^{04123}-r^5H'\Gamma^{56789}\right)=\frac{H'}{8H^{\frac{5}{4}}}\left[\Gamma^{\hat{0}\hat{4}\hat{1}\hat{2}\hat{3}}-\Gamma^{\hat{5}\hat{6}\hat{7}\hat{8}\hat{9}}\right] \ .
\end{equation}
Note that the angular part of the Dirac operator can be decomposed into the standard Dirac operator on $S^5$, together with additional contributions originating from the non-trivial embedding of $S^5$ into the six-dimensional warped geometry
\begin{equation}
    \Gamma^a D_a \Lambda =H^{-\frac{1}{4}} \Big[ \frac{1}{r} \slashed{D}_{S^5} + \frac{5}{2r} \Gamma^{\hat 4}  + \frac{1}{8} \frac{H'}{H} \Gamma^{\hat{4}} \Big] \Lambda \ ,
\end{equation}
where
\begin{equation}
    \slashed{D}_{S^5} =  \gamma^{\hat a} e^{a}_{\hat{a}} \big(\partial_a  + \frac{1}{4} \Omega_{a}^{\hat b \hat c} \Gamma_{\hat b \hat c} \big)
\end{equation}
with $\gamma^{\hat a}$  related to the 10-dimensional $\Gamma^{\hat a}$ by $\Gamma^{\hat a} = \gamma_5\otimes \gamma^{\hat a}$ in order to anti-commute with the 10-dimensional $\Gamma^{\hat\mu} = \gamma^{\hat\mu}\otimes \mathbf{1}$.
One can decompose a general ten-dimensional spinor $\Psi(X)$ in spinorial spherical harmonics 
$\widetilde{\mathcal{Y}}_{\ell, \vec{m},\eta} (\vartheta^a)$ as
\begin{equation}
\Psi(X) = \sum_{\ell,\vec{m},\eta=\pm} \psi_{\ell, 
\vec{m},\eta}(\tilde{x}) \widetilde{\mathcal{Y}}_{\ell, \vec{m},\eta} (\vartheta^a)
\end{equation}
where $\tilde{x} = (r,x^\mu)= (r,u,x^I)$ and the action of $\slashed{D}_{S^5}$ produces
\begin{equation}
    \slashed{D}_{S^5} \widetilde{\mathcal{Y}}_{\ell, \vec{m},\eta} (\vartheta^a)= \eta \widetilde\gamma_5\Gamma^{\hat{4}}  \left(\ell + \frac{5}{2} \right) \widetilde{\mathcal{Y}}_{\ell, \vec{m},\eta} (\vartheta^a)\;.
\end{equation}
Choosing a frame in which the spatial momentum along the D3-branes vanishes $p_I =0$,  the dilatino field admits the following mode expansion 
\begin{equation}\label{eq: lambda exp}
     \Lambda(u,r,\vartheta^a)=\sum_{n}\sum_{\ell, \vec{m}}\sum_{\eta}\frac{H^{\beta_{\eta}}(r)}{r^{n+5}}\Lambda^{(n)}_{\ell,\vec{m},\eta}(u)\widetilde{\mathcal{Y}}_{\ell, \vec{m},\eta} (\vartheta^a) 
\end{equation}
where $\beta_\eta$ is an $\eta$-dependent exponent to be determined momentarily.

Switching from the curved coordinates to flat-frame ones, the dilatino equation  \eqref{eq: dilatino eq} in outgoing coordinates reads
\begin{equation} \label{eq: dilatino eq 2}
   -2 H^\frac{1}{2}\mathcal{P}_+\partial_u\Lambda+ \mathfrak{D}_r \Lambda +\frac{1}{r}\Gamma^{\hat{4}}\slashed{D}_{S^5}\Lambda+\frac{H'}{8H}\left( 1 - 2 \widetilde{\gamma}_5 \right)\Lambda=0\;,
   \end{equation}
where $\mathfrak{D}_r = \partial_r + \frac{5}{2r}$ and $\mathcal{P}_\pm=\frac{1}{2}\left(1\pm\Gamma^{\hat{0}}\Gamma^{\hat{4}}\right)$ are projectors that satisfy 
\begin{equation}
    \mathcal{P}_{\pm}\widetilde{\gamma}_{5}=\widetilde{\gamma}_{5}\mathcal{P}_{\mp} \ .
    \end{equation}
Inserting the expansion \eqref{eq: lambda exp} into \eqref{eq: dilatino eq 2} yields
\begin{equation}
\sum_{n,\pm}\sum_{\ell, \vec{m}}\sum_{\eta}\Big\{\frac{2H^{1/2} }{r^{n+5}}\mathcal{P}_{+} \partial_{u}  + \frac{1}{r^{n+6}}\Big[\Big(n+\frac{5}{2}\Big)-\eta\Big(\ell + \frac{5}{2}\Big) \widetilde{\gamma}_5 \Big]  + \frac{1}{r^{n+5}}\frac{H'}{8H}\Big(1 + 8 \beta_{\eta} - 2 \widetilde{\gamma}_5 \Big)\Big\}\Lambda^{(n)\pm}_{\ell, m, \eta} \widetilde{\mathcal{Y}}_{\ell, \vec{m},\eta} = 0
\end{equation}
where $\pm$ labels eigenvalues of the operator $\Gamma^{\hat 0}\Gamma^{\hat 4}$.  
In order to obtain a quantity with $U(1)_R$ charge $+2$, we contract from the left with the 10-dimensional Killing spinor $\bar{\epsilon}^*$, which can be decomposed as \eqref{eq: killing spinor deco}. The exponents $\beta_{\eta} $ are fixed by requiring that the terms proportional to $H'/H$ cancel in each chirality sector. This condition fixes  $\beta_{\eta=1}=1/8$ and $\beta_{\eta=-1} = - 3/8$. Since we are in working in outgoing coordinates in which $\bar{\epsilon}^* (1 - \widetilde{\gamma}_5) = 0 $, only the $\eta = +1$ branch survives.

Expressing spinor spherical harmonics in terms of the scalar ones ${Y}_{\ell,  \vec{m}}$ (see Eq. \eqref{eq: spinor_spherical_harm}), projecting onto these, the last term vanishes due to the BPS projector that annihilates half of the Killing spinor components. At order $n=\ell$ we remain with
\begin{equation}
\sum_{\pm} \bar{\varepsilon}^*_{\eta=1} \Big \{ 2\sum_{k=0}^{\lfloor \frac{\ell+1}{4} \rfloor}\binom{\frac{1}{2}}{k}L^{4k} \mathcal{P}_{+}\partial_u{\Lambda}_{\ell, \vec{m},\eta=1}^{(\ell+1-4k)\pm} + \Big(n+\frac{5}{2} \Big) \Big( 1 -  \eta \widetilde{\gamma}_5 \Big) \Lambda^{(\ell)\pm}_{\ell, \vec{m},\eta=1} \Big \} =0 \ .
\end{equation}
The second term also vanishes and we obtain the conservation of the Newman-Penrose charges associated to the dilatino,
\begin{equation}
    \partial_{u}N^{\Lambda,+}_{\ell, \vec{m},\eta=1} = 0 \ ,
\end{equation}
where we identify
\begin{equation}
    N^{\Lambda, +}_{\ell,\vec{m}, \eta=1} = \sum_{k=0}^{\lfloor \frac{\ell+1}{4} \rfloor}\binom{\frac{1}{2}}{k}L^{4k} \bar{\varepsilon}^*_{\eta=1} {\Lambda}_{\ell, \vec{m},\eta= 1}^{(\ell+1-4k),+} \ .
\end{equation}

To find Aretakis charges, we consider \eqref{eq: dilatino eq} in the ingoing metric \eqref{eq: d3 hor}. Consider the following expansion for $\widetilde{\Lambda}$:
\begin{equation}
    \widetilde{\Lambda}{(v,r, \vartheta^a)}= \sum_{n,\ell,\vec{m}, \pm, \eta} r^{n} \widetilde{\Lambda}_{ \ell, \vec{m},\pm, \eta}^{(n)} (v)\mathcal{Y}_{\ell, \vec{m}, \pm, \eta}(\vartheta^a)H^{\alpha_\eta} \ ,
\end{equation} 
where $\alpha_{\eta=1} = - 3/8$ and $\alpha_{\eta=-1} = 1/8$.

We contract \eqref{eq: dilatino eq} by $\bar{\epsilon}^*$ which in these coordinates satisfies $\bar{\epsilon}^*(1 + \widetilde{\gamma}_5 ) = 0$, so only the $\eta=-1$ branch survives.  Projecting along $Y_{\ell, \vec{m}}$, we obtain a set of conserved charges in near-horizon:
\begin{equation}
    \partial_v A^{\Lambda}_{\ell, \vec{m},+, \eta=-1} = 0 \ ,
\end{equation}
where
\begin{equation}
    A^{\Lambda}_{\ell, \vec{m}, +,\eta= -1} = \sum_{k=0}^{\lfloor \frac{\ell +1}{4} \rfloor}\binom{\frac{1}{2}}{k} L^{2 - 4k} \bar{\varepsilon}^*_{\eta = -1} \widetilde{\Lambda}_{ \ell, \vec{m}, +, \eta = -1}^{(\ell + 1 - 4k)} \ .
\end{equation}

In order to check the matching between NP and Aretakis charges, we consider the conformal transformation between $\bar{\epsilon}^*\widetilde{\Lambda}$ in the near-horizon and $\bar{\epsilon}^* \Lambda$ in the near-infinity region \begin{equation}
    \bar{\epsilon}^*\widetilde{\Lambda} (v, r, \vartheta^a) = \Big( \frac{L}{r} \Big)^{5} \bar{\epsilon}^* \Lambda(u \to v, r \to L^2/r, \vartheta^a) \ .
\end{equation}
The relation between near-infinity and near-horizon modes is
\begin{equation}
    \bar{\epsilon}_0^* \widetilde{\Lambda}_{\ell, \vec{m}, \pm,\eta= -1}^{(n)} = \frac{1}{L^{2n + 5}} \bar{\epsilon}_{0}^* \Lambda^{(n)}_{\ell, \vec{m}, \pm, \eta= 1}
\end{equation}
 and consequently the Aretakis charges are proportional to the NP charges
\begin{equation}
    A^{\Lambda}_{\ell, \vec{m}, +, \eta = -1}= \frac{1}{L^{2\ell + 5}} N^{\Lambda}_{\ell, \vec{m},+, \eta = 1} \ .
\end{equation}
This reflects the fact that CT inversion flips the chirality of $\widetilde{\gamma}_5$ and therefore exchanges the two $\eta$ branches of dilatino field. Note that the relation is modified in comparison with the scalar charges as can be seen by comparing this equation with \eqref{eq:relation NP Aretakis scalar D3}.

As outlined in the previous subsection, the analysis above can be applied to other fields in the supergravity multiplet. In particular, as a natural extension of this work, one can study the linearized field equations of the doublet $B_{MN}^\alpha$ of  2-form potentials. The expansion in terms of spherical harmonics is given in subsection \ref{NPASuperfield} for the various components of the field. In particular, from such an expansion one can appreciate that there is no mixing with other components of the supermultiplet. The gauge invariance of the conserved charges implies that only the 3-form field strength will appear in the final expressions. 
A complete in-depth analysis of the higher component charges is beyond the scope of the current investigation. That said, we do not expect any major conceptual obstacle in deriving the infinite tower of conserved asymptotic charges associated to the fluctuations of the doublet of 2-forms, as well as to the ones of all the other components in the supergravity multiplet, and we hope to report soon in this direction.   

\section{Conclusions and outlook}\label{conclusions}

This work exploits Couch-Torrence conformal inversion for (intersecting) D3-brane configurations in Type-IIB supergravity in order to identify the map between Newman-Penrose charges at null infinity and Aretakis charges in the near-horizon. Both sets of charges are exactly conserved at the linear level. It is tempting to speculate a holographic interpretation in terms of protected KK modes of the supergravity fields residing in 1/2 BPS multiplets in the dual ${\cal N}=4$ SYM theory. This may provide hints towards a possible flat Carrollian holographic interpretation of NP charges at null infinity.  Supersymmetry is used to generate asymptotic charges for the higher components of the Type-IIB supergravity multiplet. In the case of the dilatino, the asymptotic charges are constructed explicitly and shown to be consistent with the supersymmetrisation of the scalar charges.

A strong motivation for our work is to find a connection between the flat-space holography program and the standard AdS/CFT picture. In this light, the investigation presented here establishes a map between the flat space and the near-horizon, $AdS_5\times S^5$ region of D3-branes, as well as their bound states. It would be very interesting to combine this bulk perspective with some of the recent developments employing a similar philosophy on the dual field theory. For instance, in the context of ABJM theory \cite{Aharony:2008ug } a recent paper \cite{Bagchi:2026emg} has discussed different ways of defining Carrollian fermions showing that the flat space limit of the bulk corresponds to the limit whereby the speed of light $c$ goes to zero in the boundary. The role of BMSvB has also been emphasized in their analysis.

Let us collect some interesting open questions for the future. A natural extension is to repeat the analysis of the dilatino equations, presented here for single D3-brane stacks to the intersecting D3-brane configurations discussed in sections \ref{sec:D3-D3} and \ref{FourD3STU}.  Similarly, one can determine the asymptotic charges associated to other fields in the supergravity multiplet, and as already mentioned the natural candidate to start this analysis would be the 2-form potential in a single D3-brane background.
It would also be interesting to go beyond the linearised level. Non-linear effects may spoil the analysis but the robustness of multiplet shortening may come to rescue. We plan to study this and related issues in the near future.

More qualitatively, one might ask what are the conditions under which brane solutions in supergravity theories can possess a CT inversion symmetry. 
For instance, D$p$-brane solutions for $p\neq 3$, do not seem to enjoy this property. Note however, that T-dual solutions to the two-stack D3-branes discussed in this work, like the M2-M5 intersection in M-theory are known \cite{Tseytlin:1996bh}, and may be a natural setting to ask such questions. More importantly, 
while in the present investigation we have emphasized  the map between null infinity and extremal non-expanding horizons, as in the original paper \cite{Couch:1984} and in the recent revival \cite{Cvetic:2020kwf, Bianchi:2021yqs, Bianchi:2022wku, Agrawal:2025fsv}, the role of the photon-sphere as a fixed-point set of \emph{generalized} CT inversions, first pointed out in \cite{Bianchi:2021yqs, Bianchi:2022wku}, was recently elaborated on in \cite{Charalambous:2025hlc}. In particular, the special case of a non-extremal `lukewarm' BH  with $M=|Q|$ in thermal equilibrium with the ambient de Sitter space was studied in great detail in \cite{Charalambous:2025hlc}.

The near extremal limit \cite{Gralla:2017lto, Gralla:2018xzo, Gralla:2019isj} and the relation between Aretakis charges and conformal invariance is another natural direction of investigation. 
Finally, it would be very interesting to consider strings, or other extended objects, propagating in a background enjoying CT inversions, and extend the notion of asymptotic charges for extended probes. 

\section*{Acknowledgments}
We thank S. Agrawal, P. Charalambous, G. Dibitetto, G. Di Russo, L. Donnay, M. Firrotta, S.M. Hosseini, E. Kiritsis, E. Marchetto, J.F. Morales, P. Pani, L. Romano, S. Rota, R. Savelli, J. Sonnenschein and D. Weissman for discussions at various stages of this work. The research of MA and MB is supported in part by an INFN fellowship under the “Iniziativa Specifica” ST\&FI.

\begin{appendix}
    \section{Frame fields, spin connections and Dirac matrices}\label{appendixA}
    \paragraph{Conventions.} 10-dimensional curved indices are denoted by upper-case latin alphabet, while under the decomposition 10=4+6, lower-case greek (latin) indices denote the 4 (6) dimensional coordinated respectively: $M=(\mu,i)$. We use a similar convention for frame indices, with the addition of a hat over the alphabet: $\widehat{M}= \left(\hat \mu, \hat i \right)$. Explicitly, these coordinates are expressed by numbers from zero to nine such that $\hat \mu =  \hat 0,  \hat 1, \hat 2, \hat 3$ and $\hat i = \hat 4, \hat 5, \dots,  \hat 9$.
    
    Consider the D3-brane metric in Cartesian coordinates
    \begin{equation}
        d s^2=H^{-\frac{1}{2}}\eta_{\mu\nu}d x^{\mu}d x^{\nu}+H^{\frac{1}{2}}\delta_{ij}dy^idy^j\;,
    \end{equation}
    where
    \begin{equation}
        H=1+\frac{L^4}{r^4}\;,\quad r=\sqrt{\delta_{ij}y^iy^j}
    \end{equation}
    From this metric one finds the following non-vanishing Christoffel symbols
    \begin{equation}
    \begin{gathered}
        \Gamma^\mu_{\nu i}=-\frac{1}{4}\delta^\mu_\nu \partial_i\log H\;,\quad \Gamma^i_{\mu\nu}=\frac{1}{4H^2}\eta_{\mu\nu}\partial^iH\;,\\\Gamma^i_{jk}=\frac{1}{4}\left(\delta^i_j\partial_k\log H+\delta^i_k\partial_j\log H-\delta_{jk}\partial^i\log H\right)
        \end{gathered}
    \end{equation}
    Define the frame fields
    \begin{equation}
        E^{\hat{\mu}}{}_\mu=H^{-\frac{1}{4}}\delta^{\hat{\mu}}{}_\mu\;,\quad E^{\hat{i}}{}_i=H^{\frac{1}{4}}\delta^{\hat{i}}{}_i{}\;.
    \end{equation}
    The non-zero spin connection components are:
    \begin{align}
        {\Omega_{\mu}}^{\hat{\nu}}{}_{\hat{i}}&=\frac{1}{4\sqrt{H}}\partial_{j}\log H \delta^{j}_{\hat i} \delta^{\hat{\nu}}_{\mu} \\
        \Omega_i{}^{\hat{k}}{}_{\hat{j}}&= \frac{1}{4} \partial_l \log H ( \delta_{i \hat{j}}\delta^{l\hat{k}}-\delta_{i}^{\hat{k}}\delta^{l}_{\hat{j}}) 
    \end{align}
\subsection{Clifford algebra}
The 10d gamma matrices are defined as a set of 32$\times$32 matrices satisfying the Clifford algebra
\begin{equation}
    \left\{\Gamma^M,\Gamma^N\right\}=2G^{MN}\;.
\end{equation}
They can be obtained from the flat space gamma matrices by a linear transformation implemented by the frame fields
\begin{equation}
\Gamma^M=E^M_{\hat{M}}\Gamma^{\hat{M}}\;,
\end{equation}
defined analogously by the flat Minkowski metric
\begin{equation}
    \left\{\Gamma^{\hat{M}},\Gamma^{\hat{N}}\right\}=2\eta^{\hat{M}\hat{N}}
\end{equation}
    We adopt a representation of the ten-dimensional gamma matrices in flat spacetime in which they factorize as tensor products of the four-dimensional and six-dimensional gamma matrices
    \begin{align}
        \Gamma^{\hat{\mu}} &= \gamma^{\hat{\mu}} \otimes I \\
        \Gamma^{\hat{i}} &= \gamma_5 \otimes \widetilde \gamma^{\hat{i}}
    \end{align}
    where $\gamma_5 = i \gamma^0\gamma^1\gamma^2\gamma^3$. The ten-dimensional chirality matrix can be written as:
    \begin{equation}
        \Gamma_{11} = \widetilde{\gamma}_5 \widetilde{\gamma}_7 = \Gamma_{\hat 0}\Gamma_{\hat 1}\Gamma_{\hat 2}\Gamma_{\hat 3}\Gamma_{\hat 4}\Gamma_{\hat 5}\Gamma_{\hat 6}\Gamma_{\hat 7}\Gamma_{\hat 8}\Gamma_{\hat 9}
    \end{equation}
    where $\widetilde{\gamma}_5 = i \Gamma_{ 0}\Gamma_{ 1}\Gamma_{ 2}\Gamma_{ 3}$ and $\widetilde{\gamma}_7 = -i \Gamma_{ 4}\Gamma_{5}\Gamma_{ 6}\Gamma_{7}\Gamma_{ 8}\Gamma_{9}$.
    \subsection{Tortoise coordinates}
    When discussing the flat space limit of the background near infinity, it is convenient to use the metric \eqref{eq: d3 inf}. The following contains various geometric quantities for this coordinate system for the reader's convenience. We decompose coordinates as follow: $X^M = (x^\mu, r,\vartheta^a)$, where $x^\mu = (u, x^I)$. The index $I=1,2,3$ labels the three longitudinal directions along the brane worldvolume, while the index $a$ denotes to the angular coordinates on $S^5$

    For the $( \hat 0, \hat 4)$ coordinates, frame fields do not take a diagonal form. Expressed as a 1-form one can choose
    \begin{equation}
        E^{\hat 0}=H^{-\frac{1}{4}}du + H^{\frac{1}{4}}dr,\quad E^{ \hat 4}=H^{\frac{1}{4}}dr
    \end{equation}
The inverse frame fields are
\begin{equation}
\begin{aligned}
E_{\hat 0} &= H^{\frac{1}{4}}\,\partial_u ,\\[4pt]
E_{\hat 4} &= -\,H^{\frac{1}{4}}\,\partial_u + H^{-\frac{1}{4}}\,\partial_r .
\end{aligned}
\end{equation}
    The angular components are:
    \begin{equation}
        E^{\hat a} = rH^{\frac{1}{4}}e^{\hat a}
    \end{equation}
    where $\hat{a}=\hat 5, \dots,\hat 9$ and $\hat e^a$ are the vielbein components for $S^5$. The frame fields along the three longitudinal directions are:
    \begin{equation}
        E^{\hat I} = H^{-\frac{1}{4}}dx^I 
    \end{equation}
    where $\hat{I}=\hat 1, \hat 2, \hat 3$.
    The metric \eqref{eq: d3 inf} is obtained via
    \begin{equation}
        ds^2 = \eta_{\hat M \hat N}E^{\hat M} E^{\hat N}
    \end{equation}
    where $\eta_{A  B}$ is the flat metric in the signature $(-,+,...,
+)$. 
    The 5-form flux in this coordinate system is 
    \begin{equation}
        F_5=H'\left[\frac{1}{H^2}d u\wedge d r\wedge d^3x-r^5d\Omega_5\right]
    \end{equation}
    In addition to the standard spin connection on $S^5$ $\Omega_a^{\hat b \hat c}$, the non-vanishing spin connection components are
    \begin{align}
        &{\Omega_u}^{\hat 0 \hat 4}{}=-{\Omega_u}^{\hat 4 \hat 0}=-\frac{H'}{4H^{3/2}} \\
        &{\Omega_r}^{\hat 0 \hat 4}=-{\Omega_r}^{\hat 4 \hat 0}= -\frac{H'}{4H} \\
        &{\Omega_I}^{\hat 4 \hat{I}}=-{\Omega_I}^{\hat{I} \hat 4}=\frac{H'}{4H^{3/2}} \\
        &{\Omega_a}^{\hat 4 \hat{a}} = - {\Omega_a}^{\hat{a} \hat 4} =  - e_a^{\hat{a}} \frac{1}{H}
    \end{align}

In near-horizon geometry, described by \eqref{eq: d3 hor}, we choose the following frame fields for the $(\hat 0, \hat 4)$ coordinates:
\begin{equation}
    E^{\hat 0} = H^{-\frac{1}{4}}dv - H^{\frac{1}{4}} dr \qquad 
    E^{\hat 4} = -H^{\frac{1}{4}}dr
\end{equation}
The inverse frame fields are:
\begin{equation}
    E_{\hat 0} = H^{\frac{1}{4}}\partial_v \qquad 
    E_{\hat 4} = -H^{\frac{1}{4}}\partial_v - H^{-\frac{1}{4}} \partial_r
\end{equation}
The angular components are:
\begin{equation}
    E^{\hat a} = - r H^{\frac{1}{4}} e^{\hat a}
\end{equation}
while frame fields along longitudinal directions are unchanged. 
The non-vanishing components of the spin connection are
\begin{align}
    &\Omega_{v}^{\hat 0 \hat 4} = - \Omega_{v}^{\hat 4 \hat 0} = \frac{H'}{4 H^{3/2}} \\
    &\Omega_{r}^{\hat 0 \hat 4}=-\Omega_{r}^{\hat 4 \hat 0} = - \frac{H'}{4H} \\ 
    &\Omega_{I}^{\hat 4 \hat{I}} = - \Omega_{I}^{\hat{I} \hat 4} = - \frac{H'}{4H^{3/2}} \\
    &{\Omega_a}^{\hat 4 \hat{a}} = - {\Omega_a}^{\hat{a} \hat 4} =   -e_a^{\hat{a}} \frac{1}{H}\;.
\end{align}

\section{KK modes of Type IIB on $S^5$ }\label{Appendix: KK modes}
Use $X^M=(x^{\tilde\mu}, \vartheta^a)$ with $x^{\tilde\mu}= (x^\mu,r) = (u, x^I, r)$ coordinates on the D3-brane plus radial direction and $\vartheta^a$ coordinates on $S^5$. The analysis is largely borrowed from \cite{Kim:1985ez}. One would naively expect a complication to arise as a result of the mixing of the metric and RR fluxes. However, neutrality of the graviton and the 5-form under $U(1)_R$ leads to a simpler picture at the linearised level. 
\subsection{Fermions}

\subsubsection{Dilatino}
We decompose the dilatino as
\begin{equation}
\Lambda(x,\vartheta) = \sum_{\ell_4,\eta} \Lambda_{\ell_4, \eta}(x) {\cal Y}_{\ell_4, \eta}(\vartheta)
\end{equation}
with ${\cal Y}_{\ell_4}^\pm$ spinor spherical harmonics, such that 
\begin{equation}
    i \slashed{D}{\cal Y}_{\ell_4, \eta } =  \eta \left(\ell_4+{5\over 2}\right) {\cal Y}_{\ell_4, \eta} \ ,
\end{equation}
and $\eta = \pm 1 $. The relation between spinor spherical harmonics and scalar ones is the following:
\begin{equation}\label{eq: spinor_spherical_harm}
    {\cal Y}_{\ell, \eta} = (i \slashed{D} + \ell +4)Y_{\ell} \kappa_{\eta} \ ,
\end{equation}
or alternatively
\begin{equation}
    {\cal Y}_{\ell, -} = (i \slashed{D} + \ell +1)Y_{\ell+1}\kappa_+ \ ,
\end{equation}
where $\kappa_{\eta}$ is the Killing spinor on $S^5$ satisfying 
\begin{equation}\label{eq: k s5}
    D_a \kappa_{\eta} + \eta \frac{i}{2} \widetilde{\gamma}_a \kappa_{\eta} = 0 \ .
\end{equation}

There are two (complex) branches of modes with 
\begin{equation}
    M_{\eta}L=\eta \left(\ell_4 + {5\over2} + \eta \right) 
\end{equation}in conjugate representations of $SU(4)$  $i.e.$  $[1,\ell_4,0]$ and $[0,\ell_4,1]$ and scaling dimension
\begin{equation}
    \Delta_{\eta} = |M_{\eta} L +  2 \eta| \ . 
\end{equation}

\subsubsection{Gravitino}
The transverse (to the sphere), $\gamma$-traceless components of gravitino can be decomposed as
\begin{equation}
    \Psi_{\tilde\mu} (x,\vartheta)= \sum_{\ell_4,\eta} \psi_{\tilde\mu, \ell_4. \eta}(x) {\cal Y}_{\ell_4, \eta}(\vartheta) \ .
\end{equation} 
Also for the gravitino there are two branches, with
\begin{equation}
    M_{\eta}L=\eta \left( \ell_4 + {5\over2} - \eta \right) \ , 
\end{equation}
in conjugate representations of $SU(4)$  $i.e.$  $[0,\ell_4,1]$ and $[1,\ell_4,0]$ and scaling dimension
\begin{equation}
    \Delta_\eta = |M_\eta L| +2 \ , 
\end{equation}

The longitudinal (to the sphere), $\gamma$-traceless components of the gravitino are 
\begin{equation}
    \Psi_{a} (x,\vartheta) = \sum_{\ell_t,\eta} \psi_{\ell_t. \eta}(x) {\cal Y}_{a, \ell_t, \eta}(\vartheta)+ \sum_{\ell_d,\eta} \psi_{\ell_d, \eta}(x) D_a{\cal Y}_{ \ell_d, \eta}(\vartheta) \ , 
\end{equation} 
where ${\cal Y}_{a, \ell_t, \eta}^\pm$ are spinor-vector spherical harmonics. Mass branches are
\begin{equation}
    M_{t,\eta}L= \eta \left( \ell_t + {5\over2} - 2 \right) \ , 
\end{equation}
in conjugate representations of $SU(4)$  $i.e.$  $36, 140,... $ and $36^*, 140^*,.. $ and 
\begin{equation}
    M_{d,\eta}L=\eta \left (\ell_t + {5\over2} -  2 \eta \right) \ ,
\end{equation}
in conjugate representations of $SU(4)$  $i.e.$  $[1,\ell_4,0]$ and $[0,\ell_4,1]$; and scaling dimension
\begin{equation}
    \Delta_\eta = |M_{d,\eta} L| + 2  \ .
\end{equation}

\subsection{Bosons}

\subsubsection{Dilaton}
The dilaton can be expanded as
\begin{equation}
    \Phi(x,\vartheta) = \sum_{\ell_1} \varphi_{\ell_1}(x) {Y}_{\ell_1}(\vartheta) \ ,
\end{equation}
with ${Y}_{\ell_1}$ scalar spherical harmonics, such that
\begin{equation}
    \nabla^2 {Y}_{\ell_1} = - \ell_1(\ell_1+4){Y}_{\ell_1} \ .
\end{equation}
There is only one (complex) branch of modes with
\begin{equation}
    M^2L^2=\ell_1(\ell_1+4) \ , 
\end{equation}
$i.e. \ \Delta=\ell_1$, where we use $\ell_1$ to label singlet of $SO(5)$.

\subsubsection{Complex 2-form / Kalb-Ramond}

The transverse components of the complex 2-form are
\begin{equation}
    B_{\tilde\mu\tilde\nu}(x,\vartheta) = \sum_{\ell_1} b_{\tilde\mu\tilde\nu}^{\ell_1}(x) {Y}_{\ell_1}(\vartheta)
\end{equation}
with ${Y}_{\ell_1}$ scalar spherical harmonics. 
There are two (complex) branches of modes: 
\begin{equation}
    M^2L^2=\ell_1^2 \qquad \ell_1 \ge 1 \ ,
\end{equation} 
in the $[0,\ell_1,0]$ of $SU(4)$ and

\begin{equation}
    M^2L^2=(\ell_1+4)^2 \qquad \ell_1 \ge 0 \ , 
\end{equation}
in the $[0,\ell_1,0]$. 

The vector components are
\begin{equation}
    B_{\tilde\mu a}(x,\vartheta) = \sum_{\ell_5} b_{\tilde\mu}^{\ell_5}(x) {\cal Y}_{a,\ell_5}(\vartheta) + \sum_{\ell_1} \tilde{b}_{\tilde\mu}^{\ell_1}(x) D_a{ Y}_{\ell_1}(\vartheta) \ . 
\end{equation}
The latter term can be gauged away, the former yield two (complex) branches of vector modes with
\begin{equation}
    M^2L^2=(\ell_5+2) (\ell_5+4)
\end{equation}
in the $[1,\ell_5,1]$ of $SU(4)$. 

The longitudinal components can be written as
\begin{equation}
    B_{ab}(x,\vartheta) = \sum_{\ell_{10}} b^{\ell_{10}}(x) {\cal Y}_{[ab]}^{\ell_{10}}(\vartheta) + \sum_{\ell_5} \tilde{b}^{\ell_5}(x) D_{[a}{\cal Y}_{b]}^{\ell_5}(\vartheta) \ . 
\end{equation}
Also in this case, the latter term can be gauged away, the former yield two (complex) branches of scalar modes with 
\begin{equation}
    M^2L^2=(\ell_{10}+3) (\ell_{10}-1) \ , 
\end{equation}
in the $[2,\ell_{10},0]$ of $SU(4)$ and 
\begin{equation}
    M^2L^2=(\ell_{10}+7) (\ell_{10}+3) \ , 
\end{equation}
in the $[2,\ell_{10},0]$ of $SU(4)$ and their complex conjugate.

\subsubsection{Graviton / RR 4-form and their mixing}

The transverse traceless graviton can be expanded as
\begin{equation}
    H_{\tilde\mu\tilde\nu}(x,\vartheta) = \sum_{\ell_1} h_{\tilde\mu\tilde\nu}^{\ell_1}(x) {Y}_{\ell_1}(\vartheta) \ . 
\end{equation}
There is one branch of symmetric traceless tensor modes with
\begin{equation}
    M^2L^2=\ell_1(\ell_1+4)
\end{equation}
in the $[0,\ell_1,0]$ of $SU(4)$.

The vector components are
\begin{equation}
    H_{\tilde\mu a}(x,\vartheta) = \sum_{\ell_1} h_{\tilde\mu}^{\ell_5}(x) {\cal Y}_{a,\ell_5}(\vartheta) + \sum_{\ell_1} \tilde{h}_{\tilde\mu}^{\ell_1}(x) D_a{ Y}_{ \ell_1}(\vartheta)  \ .
\end{equation}
The latter term can be gauged away, the former mix with the following components of the RR 4-form:

\begin{equation}
   A_{\tilde\mu abc}(x,\vartheta) = \sum_{\ell_5} a_{\tilde\mu}^{\ell_5}(x) \varepsilon_{abcde} D^d{\cal Y}^e_{\ell_5}(\vartheta) \ , 
\end{equation}
and yield two branches of real vector modes with 
\begin{equation}
    M^2L^2 = \ell_5(\ell_5+2)
\end{equation}
in the $[1,\ell_5,1]$ of $SU(4)$, and 
\begin{equation}
    M^2L^2 = (\ell_5+4) (\ell_5+6)
\end{equation}
in the $[1,\ell_5,1]$ of $SU(4)$.

The longitudinal (to the sphere) graviton components are:

\begin{equation}
    H_{(ab)}(x,\vartheta) = \sum_{\ell_{14}} h^{\ell_{14}}(x) {\cal Y}_{(ab)\ell_{14}}(\vartheta) 
    + \sum_{\ell_{5}} \tilde{h}^{\ell_{5}}(x) D_{(a}{\cal Y}_{b)\ell_{5}}(\vartheta) 
     + \sum_{\ell_{1}} \tilde{\tilde{h}}^{\ell_{1}}(x) D_{(a}D_{b)}{Y}_{\ell_{1}}(\vartheta) \ . 
\end{equation}
The last two terms are gauge trivial, the symmetric traceless part yields scalars in the $[2,\ell_{14},2]$ of $SU(4)$ with mass 
\begin{equation}
    M^2L^2= (\ell_{14}+2)(\ell_{14}+6)\ . 
\end{equation}

Now we consider the mixing between $H^a{}_a$ and $A_{abcd}$. The former can be expanded as
\begin{equation}
   H^a{}_a (x,\vartheta)= \sum_{\ell_1} h^{\ell_1}(x) {Y}_{\ell_{1}}(\vartheta)
\end{equation}
while the latter is
\begin{equation}
   A_{abcd} (x,\vartheta)= \sum_{\ell_1} a^{\ell_1}(x) \varepsilon_{abcde} D^e { Y}_{\ell_{1}}(\vartheta)
\end{equation}
Eliminating $H^{\tilde\mu}{}_{\tilde\mu}$ produces two sets of  scalar modes in the $[0,\ell_1,0]$ of $SU(4)$ and masses
\begin{equation}
   M^2L^2=\ell_1(\ell_1-4) \ ,
\end{equation}
 with $\ell_1\ge 2$ or 
 \begin{equation}
     M^2L^2=(\ell_1+4)(\ell_1+8) \ , 
 \end{equation}
 valid for any $\ell_1$.

Finally the fields $A_{\tilde\mu\tilde\nu\tilde \rho\tilde\sigma}$ (scalar ${ Y}_{\ell_{1}}$) and $A_{a\tilde\mu\tilde\nu\tilde \rho}$ (vector ${\cal Y}^{\ell_{1}}_a$) can be eliminated algebraically thanks to the self-duality constraint, while 

\begin{equation}
    A_{ab\tilde\mu\tilde\nu}(x,\vartheta) = \sum_{\ell_{10}} a^{\ell_{10}}_{\tilde\mu\tilde\nu}(x) {\cal Y}_{[ab]}^{\ell_{10}}(\vartheta) + \sum_{\ell_5} \tilde{a}^{\ell_5}_{\tilde\mu\tilde\nu}(x) D_{[a}{\cal Y}_{b]}^{\ell_5}(\vartheta) \ . 
\end{equation}
Gauging away the last term produces antisymmetric tensors in the $[2,\ell_{10},0]$ and $[0,\ell_{10},2]  $ with mass 
\begin{equation}
    M^2L^2=(\ell_{10} +3)^2
\end{equation}

\end{appendix}

\bibliography{ref}

@article{Bergshoeff:2005ac,
    author = "Bergshoeff, Eric A. and de Roo, Mees and Kerstan, Sven F. and Riccioni, Fabio",
    title = "{IIB supergravity revisited}",
    eprint = "hep-th/0506013",
    archivePrefix = "arXiv",
    reportNumber = "DAMTP-2005-47, UG-05-04",
    doi = "10.1088/1126-6708/2005/08/098",
    journal = "JHEP",
    volume = "08",
    pages = "098",
    year = "2005"
}

@article{Schwarz:1983qr,
    author = "Schwarz, John H.",
    editor = "Salam, A. and Sezgin, E.",
    title = "{Covariant Field Equations of Chiral N=2 D=10 Supergravity}",
    reportNumber = "CALT-68-1016",
    doi = "10.1016/0550-3213(83)90192-X",
    journal = "Nucl. Phys. B",
    volume = "226",
    pages = "269",
    year = "1983"
}

@article{Godazgar:2017igz,
    author = "Godazgar, Hadi and Godazgar, Mahdi and Pope, C. N.",
    title = "{Aretakis Charges and Asymptotic Null Infinity}",
    eprint = "1707.09804",
    archivePrefix = "arXiv",
    primaryClass = "hep-th",
    reportNumber = "MI-TH-1761",
    doi = "10.1103/PhysRevD.96.084055",
    journal = "Phys. Rev. D",
    volume = "96",
    number = "8",
    pages = "084055",
    year = "2017"
}

@article{Aretakis:2012ei,
    author = "Aretakis, Stefanos",
    title = "{Horizon Instability of Extremal Black Holes}",
    eprint = "1206.6598",
    archivePrefix = "arXiv",
    primaryClass = "gr-qc",
    doi = "10.4310/ATMP.2015.v19.n3.a1",
    journal = "Adv. Theor. Math. Phys.",
    volume = "19",
    pages = "507--530",
    year = "2015"
}

@article{Aretakis:2011ha,
    author = "Aretakis, Stefanos",
    title = {{Stability and Instability of Extreme Reissner-Nordstr{\"o}m Black Hole Spacetimes for Linear Scalar Perturbations I}},
    eprint = "1110.2007",
    archivePrefix = "arXiv",
    primaryClass = "gr-qc",
    doi = "10.1007/s00220-011-1254-5",
    journal = "Commun. Math. Phys.",
    volume = "307",
    pages = "17--63",
    year = "2011"
}

@article{Bianchi:2021yqs,
    author = "Bianchi, Massimo and Di Russo, Giorgio",
    title = "{Turning black holes and D-branes inside out of their photon spheres}",
    eprint = "2110.09579",
    archivePrefix = "arXiv",
    primaryClass = "hep-th",
    doi = "10.1103/PhysRevD.105.126007",
    journal = "Phys. Rev. D",
    volume = "105",
    number = "12",
    pages = "126007",
    year = "2022"
}

@article{Bianchi:2022wku,
    author = "Bianchi, Massimo and Di Russo, Giorgio",
    title = "{Turning rotating D-branes and black holes inside out their photon-halo}",
    eprint = "2203.14900",
    archivePrefix = "arXiv",
    primaryClass = "hep-th",
    doi = "10.1103/PhysRevD.106.086009",
    journal = "Phys. Rev. D",
    volume = "106",
    number = "8",
    pages = "086009",
    year = "2022"
}

@article{Agrawal:2025fsv,
    author = "Agrawal, Shreyansh and Charalambous, Panagiotis and Donnay, Laura",
    title = "{Null infinity as an inverted extremal horizon: Matching an infinite set of conserved quantities for gravitational perturbations}",
    eprint = "2506.15526",
    archivePrefix = "arXiv",
    primaryClass = "hep-th",
    month = "6",
    year = "2025"
}

@article{Cvetic:2020kwf,
    author = "Cvetic, M. and Pope, C. N. and Saha, A.",
    title = "{Generalized Couch-Torrence symmetry for rotating extremal black holes in maximal supergravity}",
    eprint = "2008.04944",
    archivePrefix = "arXiv",
    primaryClass = "hep-th",
    doi = "10.1103/PhysRevD.102.086007",
    journal = "Phys. Rev. D",
    volume = "102",
    number = "8",
    pages = "086007",
    year = "2020"
}

@article{Couch:1984,
    author = "Couch, W. E. and Torrence, R. J.",
    title = "{Conformal invariance under inversion of extreme Reissner--Nordstr{\"o}m black holes}",
    doi = "10.1007/BF00763529",
    journal = "Gen. Rel. Grav.",
    volume = "16",
    pages = "789--792",
    year = "1984"
}

@article{Kalin:2019rwq,
    author = {K{\"a}lin, Gregor and Porto, Rafael A.},
    title = "{From Boundary Data to Bound States}",
    eprint = "1910.03008",
    archivePrefix = "arXiv",
    primaryClass = "hep-th",
    reportNumber = "DESY 19-167, UUITP-40/19, DESY-19-167",
    doi = "10.1007/JHEP01(2020)072",
    journal = "JHEP",
    volume = "01",
    pages = "072",
    year = "2020"
}

@article{Bondi:1962px,
    author = "Bondi, H. and van der Burg, M. G. J. and Metzner, A. W. K.",
    title = "{Gravitational waves in general relativity. 7. Waves from axisymmetric isolated systems}",
    doi = "10.1098/rspa.1962.0161",
    journal = "Proc. Roy. Soc. Lond. A",
    volume = "269",
    pages = "21--52",
    year = "1962"
}

@article{Sachs:1962zza,
    author = "Sachs, R.",
    title = "{Asymptotic symmetries in gravitational theory}",
    doi = "10.1103/PhysRev.128.2851",
    journal = "Phys. Rev.",
    volume = "128",
    pages = "2851--2864",
    year = "1962"
}

@article{Barnich:2011mi,
    author = "Barnich, Glenn and Troessaert, Cedric",
    title = "{BMS charge algebra}",
    eprint = "1106.0213",
    archivePrefix = "arXiv",
    primaryClass = "hep-th",
    reportNumber = "ULB-TH-11-10",
    doi = "10.1007/JHEP12(2011)105",
    journal = "JHEP",
    volume = "12",
    pages = "105",
    year = "2011"
}

@article{Strominger:2013jfa,
    author = "Strominger, Andrew",
    title = "{On BMS Invariance of Gravitational Scattering}",
    eprint = "1312.2229",
    archivePrefix = "arXiv",
    primaryClass = "hep-th",
    doi = "10.1007/JHEP07(2014)152",
    journal = "JHEP",
    volume = "07",
    pages = "152",
    year = "2014"
}

@book{Ortin:2015hya,
    author = "Ortin, Tomas",
    title = "{Gravity and Strings}",
    edition = "2nd ed.",
    doi = "10.1017/CBO9781139019750",
    isbn = "978-0-521-76813-9, 978-0-521-76813-9, 978-1-316-23579-9",
    publisher = "Cambridge University Press",
    series = "Cambridge Monographs on Mathematical Physics",
    month = "7",
    year = "2015"
}

@article{Donnay:2023mrd,
    author = "Donnay, Laura",
    title = "{Celestial holography: An asymptotic symmetry perspective}",
    eprint = "2310.12922",
    archivePrefix = "arXiv",
    primaryClass = "hep-th",
    doi = "10.1016/j.physrep.2024.04.003",
    journal = "Phys. Rept.",
    volume = "1073",
    pages = "1--41",
    year = "2024"
}

@article{Newman:1961qr,
    author = "Newman, Ezra and Penrose, Roger",
    title = "{An Approach to gravitational radiation by a method of spin coefficients}",
    doi = "10.1063/1.1724257",
    journal = "J. Math. Phys.",
    volume = "3",
    pages = "566--578",
    year = "1962"
}

@article{Ruzziconi:2026bix,
    author = "Ruzziconi, Romain",
    title = "{Carrollian physics and holography}",
    eprint = "2602.02644",
    archivePrefix = "arXiv",
    primaryClass = "hep-th",
    doi = "10.1016/j.physrep.2026.03.005",
    journal = "Phys. Rept.",
    volume = "1182",
    pages = "1--87",
    year = "2026"
}

@article{Bergshoeff:2020xhv,
    author = "Bergshoeff, Eric and Izquierdo, Jos{\'e} Manuel and Romano, Luca",
    title = "{Carroll versus Galilei from a Brane Perspective}",
    eprint = "2003.03062",
    archivePrefix = "arXiv",
    primaryClass = "hep-th",
    doi = "10.1007/JHEP10(2020)066",
    journal = "JHEP",
    volume = "10",
    pages = "066",
    year = "2020"
}

@article{Charalambous:2025hlc,
    author = "Charalambous, Panagiotis and Donnay, Laura and Lupsasca, Alexandru",
    title = "{Generalized Couch-Torrence inversions}",
    eprint = "2512.18664",
    archivePrefix = "arXiv",
    primaryClass = "gr-qc",
    doi = "10.1007/JHEP03(2026)239",
    journal = "JHEP",
    volume = "03",
    pages = "239",
    year = "2026"
}

@article{Gralla:2017lto,
    author = "Gralla, Samuel E. and Zimmerman, Peter",
    title = "{Critical exponents of extremal Kerr perturbations}",
    eprint = "1711.00855",
    archivePrefix = "arXiv",
    primaryClass = "gr-qc",
    doi = "10.1088/1361-6382/aab140",
    journal = "Class. Quant. Grav.",
    volume = "35",
    number = "9",
    pages = "095002",
    year = "2018"
}

@article{Tseytlin:1996bh,
    author = "Tseytlin, Arkady A.",
    title = "{Harmonic superpositions of M-branes}",
    eprint = "hep-th/9604035",
    archivePrefix = "arXiv",
    reportNumber = "IMPERIAL-TP-95-96-38",
    doi = "10.1016/0550-3213(96)00328-8",
    journal = "Nucl. Phys. B",
    volume = "475",
    pages = "149--163",
    year = "1996"
}

@article{Godazgar:2020gqd,
    author = "Godazgar, Hadi and Godazgar, Mahdi and Perry, Malcolm J.",
    title = "{Asymptotic gravitational charges}",
    eprint = "2007.01257",
    archivePrefix = "arXiv",
    primaryClass = "hep-th",
    doi = "10.1103/PhysRevLett.125.101301",
    journal = "Phys. Rev. Lett.",
    volume = "125",
    number = "10",
    pages = "101301",
    year = "2020"
}

@article{Gralla:2019isj,
    author = "Gralla, Samuel E. and Ravishankar, Arun and Zimmerman, Peter",
    title = "{Horizon Instability of the Extremal BTZ Black Hole}",
    eprint = "1911.11164",
    archivePrefix = "arXiv",
    primaryClass = "gr-qc",
    doi = "10.1007/JHEP05(2020)094",
    journal = "JHEP",
    volume = "05",
    pages = "094",
    year = "2020"
}

@article{Gralla:2018xzo,
    author = "Gralla, Samuel E. and Zimmerman, Peter",
    title = "{Scaling and Universality in Extremal Black Hole Perturbations}",
    eprint = "1804.04753",
    archivePrefix = "arXiv",
    primaryClass = "gr-qc",
    doi = "10.1007/JHEP06(2018)061",
    journal = "JHEP",
    volume = "06",
    pages = "061",
    year = "2018"
}

@article{Gunaydin:1984fk,
    author = "Gunaydin, M. and Marcus, N.",
    title = "{The Spectrum of the s**5 Compactification of the Chiral N=2, D=10 Supergravity and the Unitary Supermultiplets of U(2, 2/4)}",
    reportNumber = "CALT-68-1185",
    doi = "10.1088/0264-9381/2/2/001",
    journal = "Class. Quant. Grav.",
    volume = "2",
    pages = "L11",
    year = "1985"
}

@article{Kim:1985ez,
    author = "Kim, H. J. and Romans, L. J. and van Nieuwenhuizen, P.",
    title = "{The Mass Spectrum of Chiral N=2 D=10 Supergravity on S**5}",
    reportNumber = "ITP-SB-84-87",
    doi = "10.1103/PhysRevD.32.389",
    journal = "Phys. Rev. D",
    volume = "32",
    pages = "389",
    year = "1985"
}

@article{Bagchi:2026emg,
    author = "Bagchi, Arjun and Lipstein, Arthur and Mondal, Saikat and Zhang, Alex Jiayi",
    title = "{Carrollian ABJM: Fermions and Supersymmetry}",
    eprint = "2604.22582",
    archivePrefix = "arXiv",
    primaryClass = "hep-th",
    month = "4",
    year = "2026"
}

@article{Aharony:2008ug,
    author = "Aharony, Ofer and Bergman, Oren and Jafferis, Daniel Louis and Maldacena, Juan",
    title = "{N=6 superconformal Chern-Simons-matter theories, M2-branes and their gravity duals}",
    eprint = "0806.1218",
    archivePrefix = "arXiv",
    primaryClass = "hep-th",
    reportNumber = "WIS-12-08-JUN-DPP",
    doi = "10.1088/1126-6708/2008/10/091",
    journal = "JHEP",
    volume = "10",
    pages = "091",
    year = "2008"
}

@article{Chen:2025sim,
    author = "Chen, Calvin Y. -R. and Kov{\'a}cs, {\'A}ron D.",
    title = "{On the Aretakis instability of extremal black branes}",
    eprint = "2507.12529",
    archivePrefix = "arXiv",
    primaryClass = "hep-th",
    doi = "10.1007/JHEP09(2025)084",
    journal = "JHEP",
    volume = "09",
    pages = "084",
    year = "2025"
}
\end{document}